\documentclass[11pt,a4paper]{article}

\usepackage{bm,amsfonts,amssymb,mathrsfs,amsmath}
\usepackage{graphicx}
\usepackage{hyperref}
\usepackage{ulem}
\usepackage{color}

\newcommand{\bZ}{\mathbb{Z}}

\newcommand{\bea}{\begin{eqnarray}}
\newcommand{\eea}{\end{eqnarray}}

\makeatletter
\@addtoreset{equation}{section}
\makeatother

\date{\today}

\begin{document}

\begin{titlepage}

\renewcommand{\thefootnote}{\fnsymbol{footnote}}

\begin{flushright}
 {\tt 
 OU-HET-870
 }
\\
\end{flushright}

\vskip9em

\begin{center}
 {\LARGE
 \textbf{
 Band spectrum is D-brane 
 }}

 \vskip5em

 \setcounter{footnote}{1}
 {\sc Koji Hashimoto$^{1}$}\footnote{E-mail address: 
 \href{mailto:koji@phys.sci.osaka-u.ac.jp}
 {\tt koji(at)phys.sci.osaka-u.ac.jp}}
 and
 \setcounter{footnote}{2}
 {\sc Taro Kimura$^{2}$}\footnote{E-mail address: 
 \href{mailto:taro.kimura@keio.jp}
 {\tt taro.kimura(at)keio.jp}}

 \vskip2em

{\it 
 $^{1}$%
 Department of Physics, Osaka University,
 Toyonaka, Osaka 560-0043, Japan
 \\ \vspace{.5em}
 $^{2}$%
 Department of Physics, Keio University, Kanagawa
 223-8521, Japan
}

 \vskip3em

 \end{center}

 \vskip2em

 \begin{abstract}
We show that band spectrum of topological insulators
can be identified as the shape of D-branes in string theory.
The identification is based on a relation between the Berry connection
associated with the band structure
and the ADHM/Nahm construction of solitons
whose geometric realization is available with D-branes.
We also show that chiral and helical edge states are identified as 
D-branes representing a noncommutative monopole.
 \end{abstract}

\end{titlepage}

\tableofcontents

\hrulefill

\vspace{1em}

\setcounter{footnote}{0}

\section{Introduction}
\label{sec1}

Topological insulators and superconductors are one of the most
interesting materials in which theoretical and experimental progress
have been intertwined each other.
In particular, the classification of topological phases 
\cite{Schnyder:2008tya,Kitaev:2009mg}
provided concrete and rigorous argument on stability and possibility of
topological insulators and superconductors.
The key to find the topological materials is their electron band structure.
The existence of gapless edge states appearing at spatial boundaries
of the material signals the topological property. Identification of possible
electron band structures is directly related to the topological nature of the
topological insulators. It is important, among many possible applications
of the topological insulators, to gain insight on what kind of electron
band structure is possible for topological insulators with fixed
topological charges.

D-branes in superstring theory \cite{Dai:1989ua,Polchinski:1994fq,Polchinski:1995mt}
are extended objects in higher spatial dimensions
which play crucial roles in any string theory dynamics. The shape of D-branes
encodes information of the higher dimensions as well as non-perturbative
dynamics of gauge theories living on the D-branes. In particular, D-branes have Ramond-Ramond charges which can be seen as topological charges on the 
brane worldvolume theories. The D-brane charges are classified by K-theory \cite{Witten:1998cd}, which offers a natural path to relate the topological insulators
and superstring theory. Indeed, recent progress
\cite{Ryu:2010hc,Ryu:2010fe,Furusaki:2012zf}
realizes a field theory setups of 
the topological insulators in terms of worldvolume gauge theories on D-branes,
which provides a consistent K-theory interpretation with 
Refs.~\cite{Kitaev:2009mg,Ryu:2010zza}. The established relation is partially due to
the Chern-Simons term indicating the topological nature of the theory,
appearing  as a part of the D-brane worldvolume gauge theory.

The topological nature of the topological insulators is, on the other hand,
naturally understood in terms of electron band structure. The quantum Hall effect,
which is the most popular example of the topological material,
has the topological number called Thouless-Kohmoto-Nightingale-den Nijs (TKNN)
number \cite{Thouless:1982zz}. The topological number $\nu_{\rm TKNN}$ is defined
as a first Chern class of the Berry connection of Bloch wave functions in the 
momentum space. The electron band structure crucially determines the Chern number, and the topological nature is hidden in the band spectrum.

In this paper, we show that the electron band structure of
topological insulators
can be identified as the shape of D-branes in string theory.
The dispersion relation in the momentum space for electrons
in the continuum limit is shown to be
identical to the shape of a particular species of D-branes in higher dimensional
coordinate space. 

To relate these, we notice the following analogy between (i) the D-branes, 
(ii) topological solitons and (iii) the topological insulators.
\begin{itemize}
\item (i) $\leftrightarrow$ (ii). Particular set of D-branes in string theory can represent topological solitons
of gauge theories \cite{Witten:1995gx,Douglas:1995bn,Callan:1997kz}.
Monopoles can be given by a D1-brane stuck to D3-branes
as seen from the D3-brane worldvolume theory. The D3-brane
exhibits a particular spiky shape in higher dimensional space. Instantons can be given
by a D0-brane bound inside a pile of D4-branes.
\item (ii) $\leftrightarrow$ (iii).
For important kinds of topological solitons, there exists a construction method
of all possible solitons. For monopoles and instantons, we have Nahm construction
of monopoles \cite{Nahm:1979yw}
and Atiyah-Drinfeld-Hitchin-Manin (ADHM) construction of instantons \cite{Atiyah:1978ri,Corrigan:1983sv}. 
The way they work is quite analogous to the Berry connections in topological insulators.
\end{itemize}
We fully use these correspondence to find that the shape of D-branes in 
coordinate spaces can be identical to the shape of the electron bands in topological
insulators in momentum space.

In~\cite{Horava:2005jt}, the stability of the Fermi surfaces was topologically studied 
from the viewpoint of K-theory, and possible relation to D-branes
due to the K-theory was pointed out, 
through the exchange of the coordinate space and the momentum space. 
See also~\cite{Rey:2008zz} and~\cite{Volovik:2009univ}.
Based on this exchange, we find an explicit and new connection between
the electron bands and the shape of the D-branes.%
\footnote{So, to find an explicit relation between ours and 
Refs.~\cite{Ryu:2010hc,Ryu:2010fe,Furusaki:2012zf}
which do not use the exchange is an open question.}

We first study typical examples of class A topological insulators both in 2
and 4 dimensions, which have no additional discrete
symmetry, according to the established
classification of topological phases~\cite{Schnyder:2008tya,Kitaev:2009mg}.
We start with a popular topological property of a Hamiltonian of a free
electron, and will find that it parallels the Nahm and ADHM constructions
of monopoles and instantons.
Such topological solitons
respect supersymmetries in string theory and are represented by
a set of D-branes. 
To have a direct relation, we use the fact that 
some scalar field is associated with the 
topological solitons, through Bogomol'nyi-Prasad-Sommerfield (BPS) equations. 
The scalar field configuration is 
the shape of D-branes,
and we can show that 
it can be identified with
the electron dispersion relation
through the connection described. 

\begin{figure}[h]
\begin{center}
\includegraphics[width=6cm]{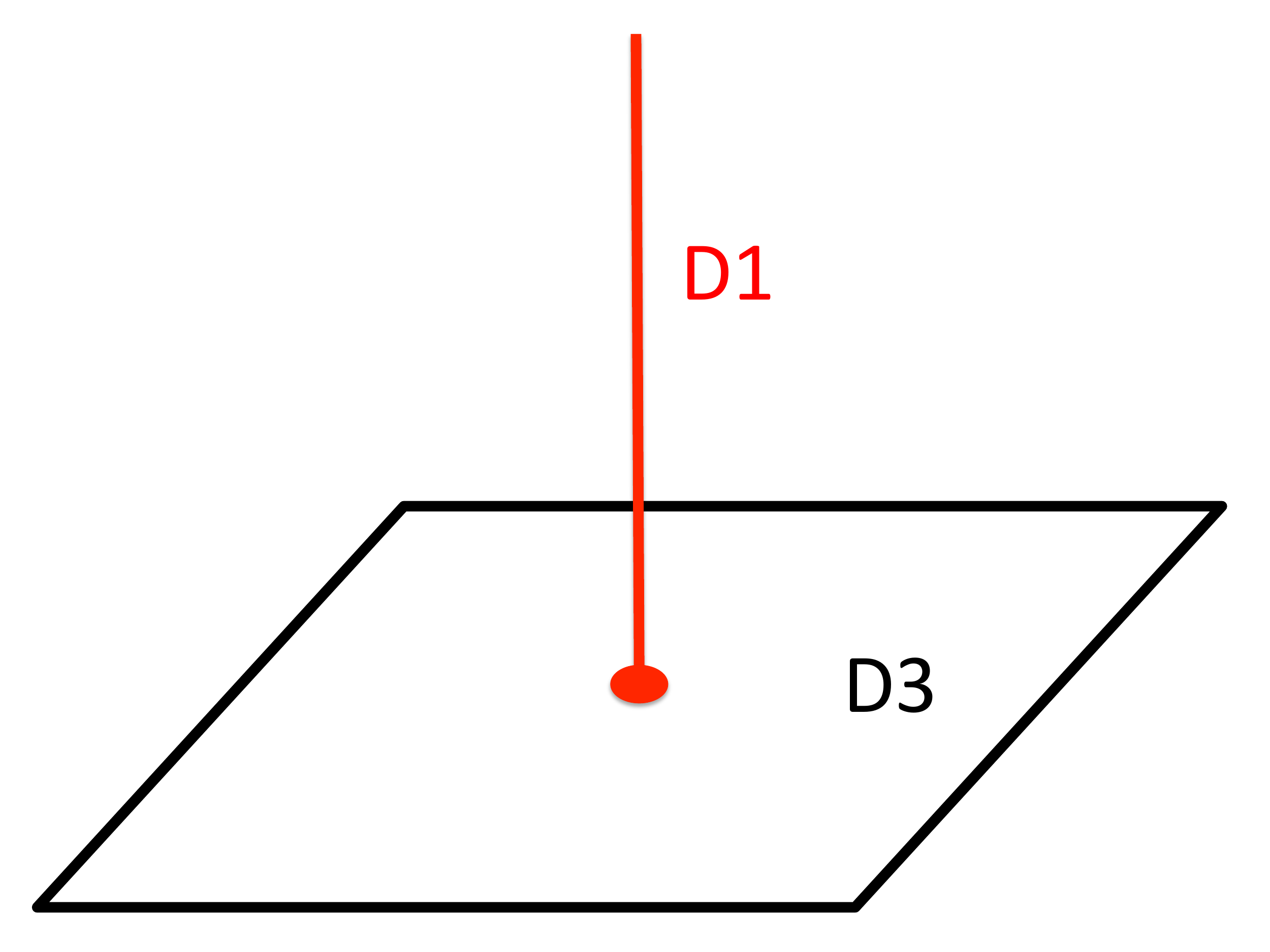}
\includegraphics[width=6cm]{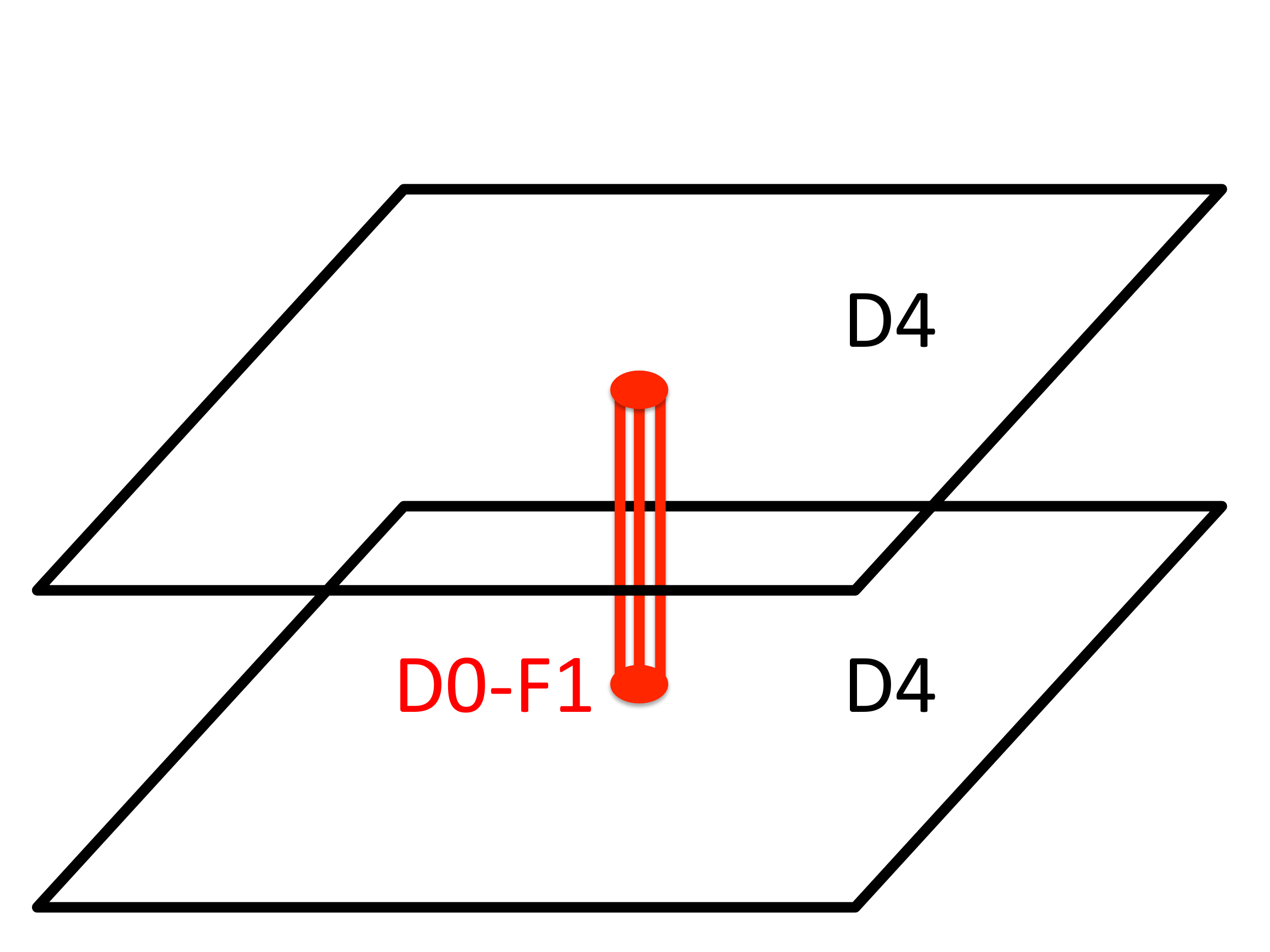}\\[5mm]
\includegraphics[width=6cm]{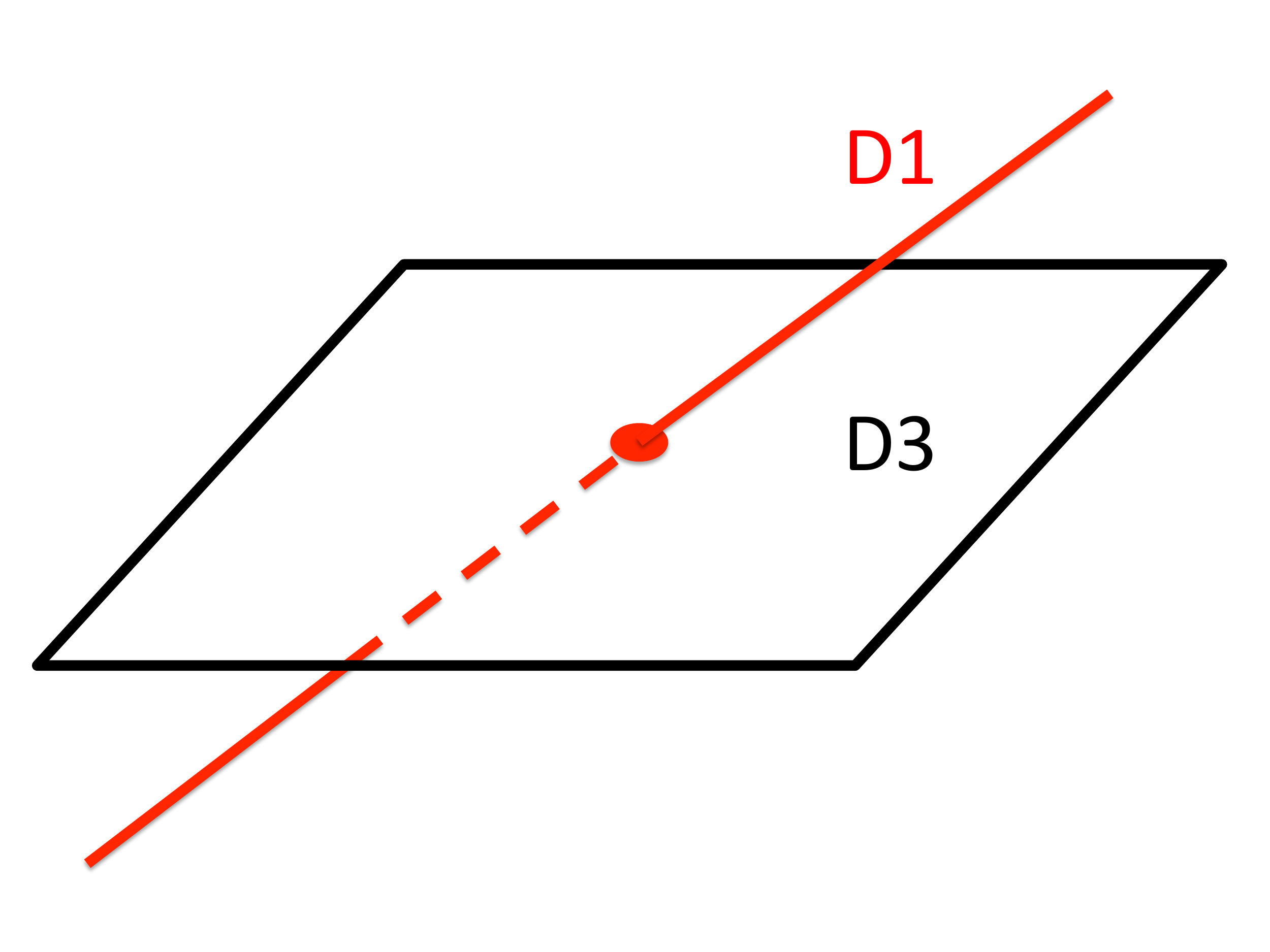}
\includegraphics[width=6cm]{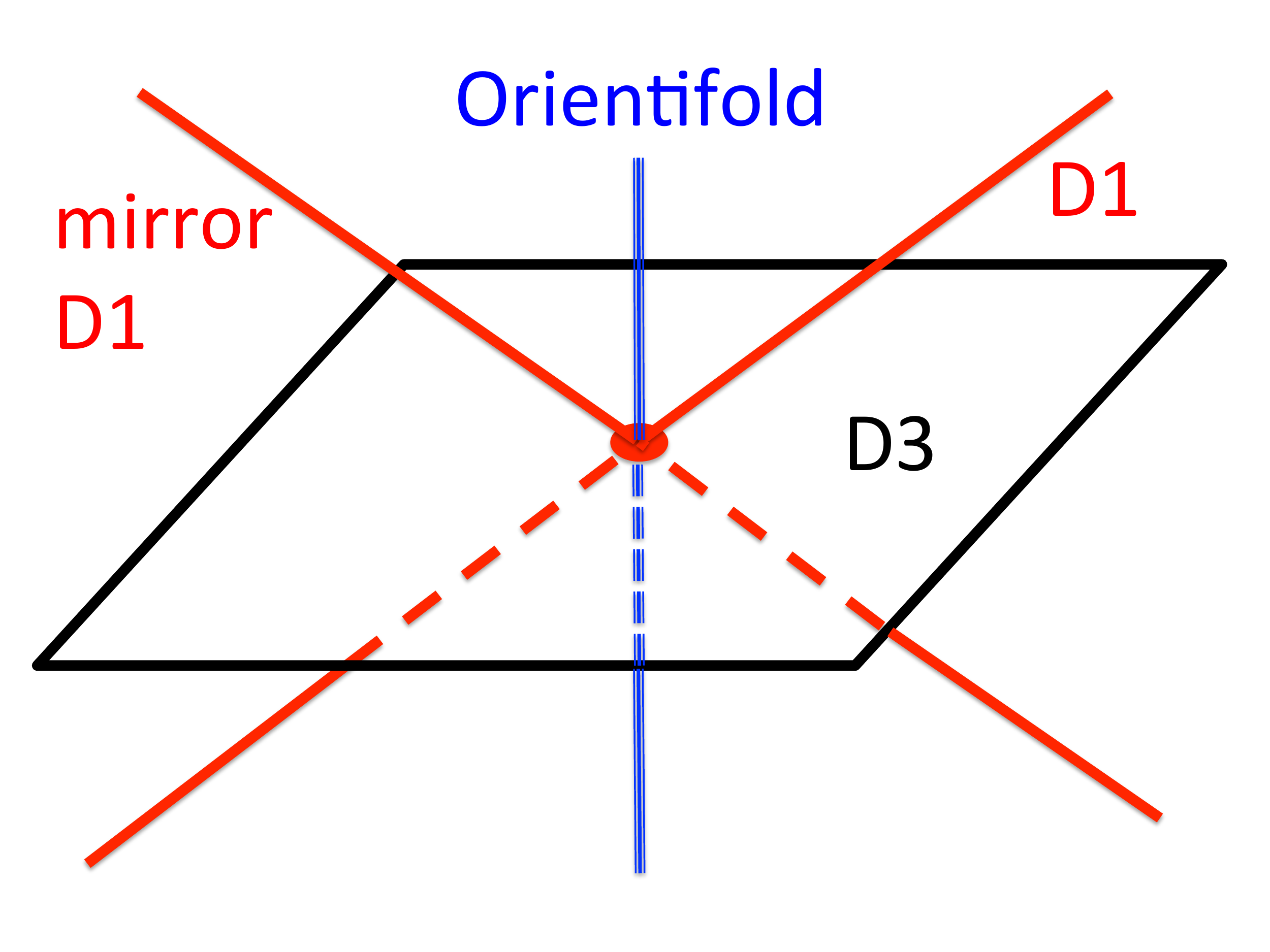}
\end{center}
\caption{Top-Left: a D1-brane stuck to a D3-brane, corresponding to a 
$U(1)$ monopole, and to an electron band of the 2-dimensional topological insulator.
Top-Right: a D0-F1(fundamental string) bound state stuck to two parallel D4-branes,
corresponding to a dyonic instanton, and to a band structure of the 4-dimensional topological insulator. Bottom-Left: a D1-brane piercing a D3-brane at an oblique angle, 
corresponding to an energy dispersion of a chiral edge state at the boundary of the 2-dimensional class A topological insulator.
Bottom-Right: An orientifold makes a mirror image of the slanted D1-brane, corresponding to an
energy dispersion of a helical edge state at the boundary of the 2-dimensional class AII topological insulator.}
\label{D-fig} 
\end{figure}

In this paper we consider the following D-brane configurations:
D1-branes stuck to a D3-brane, and a D0-brane within D4-branes
with fundamental strings stuck to them, and a D1-brane piercing a D3-brane 
at an oblique angle (see Fig.~\ref{D-fig}). Through (i) $\leftrightarrow$ (ii),
each brane setup corresponds 
respectively to $U(1)$ monopoles, dyonic instantons \cite{Lambert:1999ua}
and a monopole in non-commutative space \cite{Hashimoto:1999zw,Gross:2000wc,Gross:2000ph,Gross:2000ss}. 
Then, through (ii) $\leftrightarrow$ (iii), each corresponds 
respectively to 2-dimensional and 4-dimensional class A topological
insulators, and a chiral edge state at the boundary of the 2-dimensional
topological insulators.
In each example, we find that the D-brane shape is the
electron band structure.

This argument can be generalized to other classes by applying
discrete symmetry.
We in particular study the class AII system by imposing time-reversal
symmetry, and show how the helical edge state, which is peculiar to
this case, can be understood from the string theoretical point of view.
The symmetry applied here can be realized by using an orientifold, and
we show how it stabilizes the D-brane configuration corresponding to
the helical edge state.

The merit of presenting a map between the band structure of electrons and 
the shape of D-branes is to have a better understanding of possible band
structure, as D-branes have fertile and fruitful applications in string
theory in higher dimensions.
In this paper, we partially identify the mechanism of how topological
insulators can acquire a topological number larger than 1 for
class A systems.
We solve the constraint equation of the soliton construction techniques,
which are called the Nahm equation, to find a class of electron
Hamiltonians which have larger and generic topological numbers, and
discuss its realization in a multilayer system.

The organization of this paper is as follows.
In section \ref{sec2}, we study the 2-dimensional class A topological
insulators.
We briefly review the Nahm construction of monopoles, and study its
relevance to the Hamiltonian, Berry connection and the topological number.
Then we identify the shape of the deformed D3-branes as the electron
dispersion relation.
We study how larger topological numbers are realized in Hamiltonians, through
the shape of the D-branes.
In section \ref{sec3}, we further generalize the correspondence to the
4-dimensional class A topological insulators.
We will find that the D-brane shape is related to the electron
dispersion through the dyonic instanton which probes the shape of the
instanton through the additional scalar field.
In section \ref{sec4}, we turn to a boundary of the 2-dimensional
topological insulators.
There, the electron Hamiltonian is identified with the Nahm construction in
a noncommutative space.
The solutions of noncommutative monopoles are represented by a slanted
D1-brane which is shown to relate directly to the shape of the
dispersion relation of the chiral edge state.
In section \ref{sec5}, we generalize this argument to the class AII
system in 2 dimensions by imposing time-reversal symmetry.
We show that this symmetry is realized by using an orientifold for the
D-brane configuration.
It naturally requires a mirror D1-brane which is allowed to intersect
with the orientifold as a pair.
This pair of D1-branes exhibits the helical edge state in the class AII
topological insulator.
The final section is for a summary and
discussions for further applications.

\section{2D class A and D1-D3 brane systems}
\label{sec2}

In this section, we show that a band spectrum of a 2 dimensional class A topological insulator is identified as the shape of a D-brane.

\subsection{A brief review of Nahm construction of monopoles}

Our identification of the band and the D-brane is based on 
Nahm construction of monopoles%
\footnote{For the D-brane interpretation of the Nahm equation, see
\cite{Diaconescu:1996rk}.
For the D-brane interpretation of the Nahm construction itself, see
\cite{Hashimoto:2005yy}.}
~\cite{Nahm:1979yw}.
It 
is a complete process to construct all solutions
of BPS equations for monopoles. 
In particular,
it is useful for constructing multiple monopole solutions in non-Abelian gauge theories, but here for our purpose a single monopole in Abelian gauge theory suffices. We shall give a brief review of how it is constructed.

First we prepare for a ``Dirac operator'' $\nabla^\dagger$ in one dimension parameterized by a coordinate $\xi$, 
\begin{eqnarray}
\nabla^\dagger \equiv i \frac{d}{d\xi} + i \sigma_i \left(x^i-T_i(\xi)\right) \, .
\end{eqnarray}
Here $\sigma_i (i=1,2,3)$ is the Pauli matrix, and $T_i(\xi)$ is a $k \times k$ 
Hermitian matrix which satisfies the Nahm equation,
\begin{eqnarray}
\frac{d}{d\xi} T_i = i\epsilon_{ijk} T_j T_k\, .
\label{Nahm}
\end{eqnarray}
The matrix size $k$ is for $k$ monopoles.
For example, for a single monopole $k=1$, the Nahm equation is trivially solved by $T_i(\xi)=0$. 

Next, we solve the ``Dirac equation''
\begin{eqnarray}
\nabla^\dagger v(\xi) = 0
\label{DiracNahm}
\end{eqnarray}
where the vector $v(\xi)$ is normalized as
\begin{eqnarray}
\int_{-\infty}^0 d\xi \; v^\dagger v = 1. 
\end{eqnarray}
This expression is for $U(1)$ BPS monopoles.\footnote{For $SU(2)$ monopoles,
the integration region is chosen to a finite period, $-1 < \xi < 1$. Then there appears
two zero modes which are ortho-normalized as $\int_{-1}^1
d\xi \; (v^{(m)})^\dagger v^{(n)} = \delta_{mn}$ for $m,n=1,2$.}

Finally, the monopole solution satisfying a $U(1)$ BPS equation
\begin{eqnarray}
\partial_i \Phi(x) = \frac12 \epsilon_{ijk} F_{jk}(x)
\label{BPSeq}
\end{eqnarray}
is given by 
\begin{eqnarray}
\Phi(x) \equiv \int \!d\xi \; v^\dagger \xi v \, , 
\quad
A_i(x) \equiv  \int \! d\xi \; v^\dagger i\frac{d}{dx^i} v \, ,
\label{PhiAformula}
\end{eqnarray}
where $\Phi$ and $A_i (i=1,2,3)$ are Hermitian.

As an exercise,  using the Nahm construction 
let us construct a BPS $U(1)$ Dirac monopole solution 
\begin{eqnarray}
\Phi(x) = \frac{-1}{2r}\, , \quad B_i \left(=\frac12 \epsilon_{ijk} F_{jk}\right)
= \frac{x^i}{2r^3} \, ,
\label{Diracm}
\end{eqnarray}
where $r\equiv \sqrt{(x^1)^2 + (x^2)^2+(x^3)^2}$. The normalized 
zero mode solving the ``Dirac equation'' (\ref{DiracNahm}) is easily obtained as\footnote{Note that the other zero mode which has $\exp(-r\xi)$ instead of $\exp(r\xi)$ is not normalizable
for $-\infty < \xi < 0$. The mode would have been normalizable and 
necessary if one wanted a 
$SU(2)$ 't Hooft Polyakov monopole solution, given by $-1<\xi<1$.}
\begin{eqnarray}
v = \frac{1}{\sqrt{r + x^3}}\exp(r \xi) \; \left(
\begin{array}{c}-x^1+ix^2 \\ r + x^3\end{array}\right) \, .
\label{zerovec}
\end{eqnarray}
Then using the formulas (\ref{PhiAformula}), we obtain the BPS Dirac monopole solution (\ref{Diracm}). The Dirac string is at the negative axis of $x^3$, as seen from
the vector (\ref{zerovec}) having an ill-defined normalization factor there.

As for our later purpose let us construct $k=2$ monopole solution. The Nahm equation
(\ref{Nahm}) can be solved by
\begin{eqnarray}
T_1 = \Delta \sigma_1,\;  T_2=T_3=0
\label{T1delta}
\end{eqnarray}
where $\Delta$ is a constant parameter.\footnote{Other solution which represents
a ``fuzzy funnel'' is $T_i = \sigma_i/\xi$.} With this, the ``Dirac equation'' (\ref{DiracNahm}) is given by
\begin{eqnarray}
\frac{d}{d\xi}v + \left(
\begin{array}{cccc}
x^3 & x^1-ix^2 & & \Delta \\
x^1+ix^2 & -x^3 & \Delta & \\
& \Delta & x^3 & x^1-ix^2 \\
\Delta & & x^1+ix^2 & -x^3
\end{array}
\right) v = 0.
\end{eqnarray}
The solutions are
\begin{eqnarray}
v_{-} = \frac{\exp[r_- \xi]}{\sqrt{2(x^3+r_-)}}
\left(
\begin{array}{c}
 x^1-i x^2-\Delta \\
 -(x^3+r_{-})\\
-(x^1-i x^2-\Delta)
 \\
 x^3+r_{-}
 \end{array}
\right),
\quad
v_{+} = \frac{\exp[r_+ \xi]}{\sqrt{2(x^3+r_+)}}
\left(
\begin{array}{c}
x^1-i x^2+ \Delta
\\
- (x^3+r_{+})
\\
x^1-i x^2+\Delta
\\
- (x^3+r_{+})
\end{array}
\right),
\nonumber
\\ 
\label{v-v+}
\end{eqnarray}
where $r_\pm \equiv \sqrt{(x^1\pm\Delta)^2 + (x^2)^2+(x^3)^2}$.
These vectors are orthogonal to each other and normalized.
The scalar field is obtained by the Nahm construction formula as
\begin{eqnarray}
\Phi = \int_{-\infty}^0 d\xi \left(v_-^\dagger \xi v_- + v_+^\dagger \xi v_+\right)
= \frac{-1}{2r_-} + \frac{-1}{2r_+} \, .
\end{eqnarray}
We find that two $U(1)$ 
monopoles are located at $(x^1,x^2,x^3)=(\pm \Delta,0,0)$,
and the monopole charge is two. The Dirac strings are in the negative $x^3$ 
direction emanating from each monopole.

\subsection{The shape of D-brane relates to electron band structure}

We start with the model Hamiltonian in 2
dimensions, describing the vicinity of the band crossing point,
\begin{eqnarray}
{\cal H} = \sigma_1 p_1 + \sigma_2 p_2 + \sigma_3 m
\label{Hamil2d}
\end{eqnarray}
where $p_1, p_2$ are momentum of the electron and $m$ is the band gap,
playing a role of the mass term.
The eigenvalues of this Hamiltonian \eqref{Hamil2d} are simply given by 
\begin{eqnarray}
\epsilon = \pm \sqrt{p_1^2 + p_2^2 + m^2} \, .
\label{2deigen}
\end{eqnarray}
This is a dispersion relation of a relativistic particle with its mass $m$.

\begin{table}[t]
 \begin{center}
  \begin{tabular}{c|cccccccc|ccc}\hline\hline
   class $\backslash$ $d$ & 0 & 1 & 2 & 3 & 4 & 5 & 6 & 7 & T & C & S \\ \hline
   A & $\bZ$ & 0 & $\bZ$ & 0 & $\bZ$ & 0 & $\bZ$ & 0 & 0 & 0 & 0 \\
   AIII & 0 & $\bZ$ & 0 & $\bZ$ & 0 & $\bZ$ & 0 & $\bZ$ & 0 & 0 & 1 \\ \hline
   AI & $\bZ$ & 0 & 0 & 0 & $2\bZ$ & 0 & $\bZ_2$ & $\bZ_2$ & $+$ & 0 & 0 \\   
   BDI & $\bZ_2$ & $\bZ$ & 0 & 0 & 0 & $2\bZ$ & 0 & $\bZ_2$ & $+$ & $+$ &1 \\
   D & $\bZ_2$ & $\bZ_2$ & $\bZ$ & 0 & 0 & 0 & $2\bZ$& 0 & 0 & $+$ & 0 \\ 
   DIII & 0 & $\bZ_2$ & $\bZ_2$ & $\bZ$ & 0 & 0 & 0 & $2\bZ$ & $-$ & $+$ & 1\\
   AII & $2\bZ$ & 0 & $\bZ_2$ & $\bZ_2$ & $\bZ$ & 0 & 0 & 0 & $-$ & 0 & 0 \\
   CII & 0 & $2\bZ$ & 0 & $\bZ_2$ & $\bZ_2$ & $\bZ$ & 0 & 0 & $-$ & $-$ & 1\\ 
   C & 0 & 0 & $2\bZ$ & 0 & $\bZ_2$ & $\bZ_2$ & $\bZ$ & 0 & 0 & $-$ & 0 \\ 
   CI & 0 & 0 & 0 & $2\bZ$ & 0 & $\bZ_2$ & $\bZ_2$ & $\bZ$ & $+$ & $-$ & 1 \\ 
   \hline\hline
  \end{tabular}
  \caption{Classification of topological insulators and topological
  superconductors~\cite{Schnyder:2008tya,Kitaev:2009mg}.
  This classification is associated with discrete symmetries,
  time-reversal (T), particle-hole (C), and chiral (sublattice) (S)
  symmetries.
  A system classified into class A which we focus on in this part has
  no additional discrete symmetry.
  We will also discuss the class AII system by imposing time-reversal
  symmetry in section~\ref{sec5}.
}
  \label{PT}
 \end{center}
\end{table}

This system is classified into class A according to the periodic table
of topological insulators~\cite{Schnyder:2008tya,Kitaev:2009mg}.
As summarized in Table~\ref{PT}, the class A system has topological
charge $\bZ$ in even dimensions, and no additional discrete symmetry.
In this sense it is the most generic situation, which can be a good
starting point to study.
Other classes can be realized by imposing additional symmetries.
In section~\ref{sec5}, for example, we will explain how to incorporate the
time-reversal symmetry.
We remark that this classification is completely parallel to possible D-brane
charges based on K-theory~\cite{Witten:1998cd}, and the class A
corresponds to type IIA string theory having D-branes in even dimensions.

Let us point out a relevance to the ``Dirac equation'' (\ref{DiracNahm})
of the Nahm construction of monopoles.
We find that (\ref{DiracNahm}) for a single $U(1)$ monopole is identical
to the Hamiltonian time evolution
\begin{eqnarray}
\left[ i\frac{\partial}{\partial t} + {\cal H} \right] v(t) = 0
\end{eqnarray}
with the Hamiltonian (\ref{Hamil2d}), when one identifies
$\xi \leftrightarrow i t $ and 
\begin{eqnarray}
(x^1,x^2,x^3) \leftrightarrow (p_1,p_2,m).
\label{xp}
\end{eqnarray}

Using the eigenvalues (\ref{2deigen}), the ``Dirac operator'' $\nabla^\dagger$
is written as
\begin{eqnarray}
\nabla^\dagger \sim \frac{\partial}{\partial \xi} \pm \epsilon \, ,
\end{eqnarray}
so the zero mode of it is proportional to $\exp[\epsilon \xi]$. This particular form of
the eigenfunction, together with 
$\epsilon \sim \int \! d\xi \; v^\dagger (\partial/\partial \xi) v$, leads to
a novel relation
\begin{eqnarray}
\Phi = \int d\xi \; v^\dagger \xi v \sim \frac{1}{\epsilon}. 
\end{eqnarray}
A precise expression between the electron energy
and the scalar field of 
the BPS monopole via the Nahm construction is found as
\begin{eqnarray}
\pm 2\epsilon = \frac{1}{\Phi} \, .
\label{energyscalar}
\end{eqnarray}

The scalar field of the BPS monopole is nothing but the shape of the D-brane:
a D1-brane stuck perpendicular to a D3-brane \cite{Callan:1997kz}. The scalar field is a deformation
of the D3-brane surface, causing a spike configuration. When identifying the
electron dispersion with the shape of a D-brane, we note two points:
\begin{itemize}
\item The coordinate in which the D-brane lives is interpreted as the electron's
momentum and the mass, as in (\ref{xp}).
\item The D-brane shape is measured in a space in which the transverse coordinate
is inverse of the original flat coordinate, $X = 1/\Phi$. 
\end{itemize}
In particular, the latter changes the spike shape $\phi \sim 1/r$ 
(which would be popular in string theory) 
to a conic shape $\Phi \sim r$.
This transformation is just one of the general coordinate transformation of the
target space of string theory, and is popular for the case of $AdS$ geometries,
see for example Ref.~\cite{Drukker:2005kx}.

\begin{figure}[h]
 \includegraphics[width=7cm]{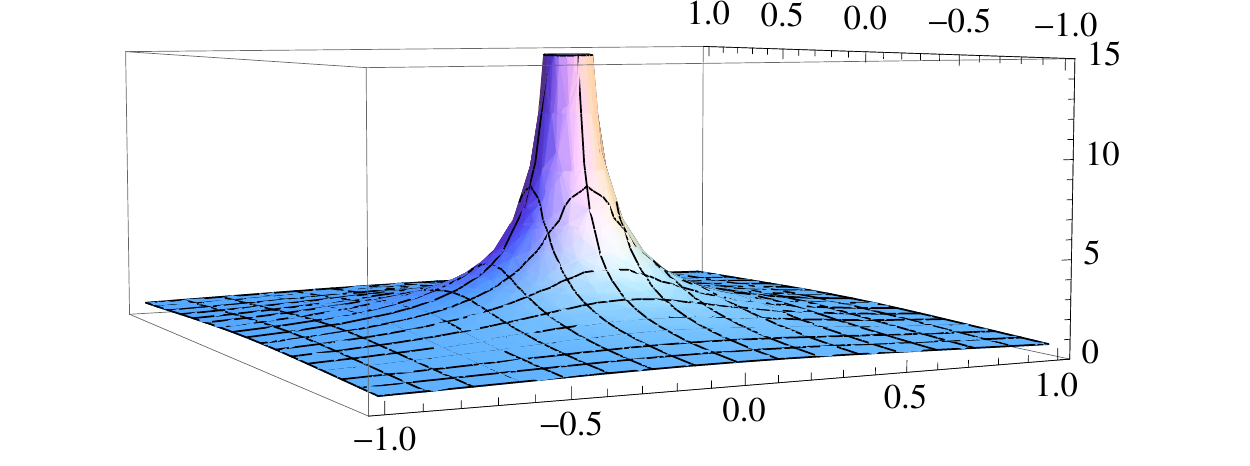}
\includegraphics[width=7cm]{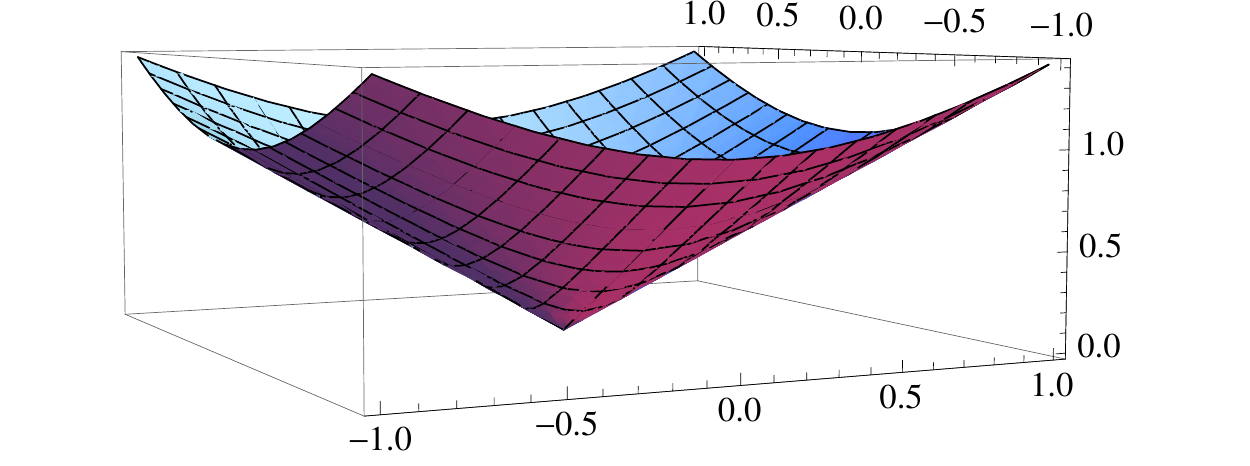}
\caption{Left: a spike configuration representing a 
D1-brane stuck to a D3-brane. The D1-brane can be
seen as a monopole from the D3-brane gauge theory.
Right: the same configuration in a different coordinate $X=1/\Phi$. 
The coordinate in
the vertical
direction is replaced by its inverse.}
\label{spike-cone}
\end{figure}

In Fig.~\ref{spike-cone}, we plot a spike configuration of the D1-D3 brane system,
in the two ways, for $m=0$. 
The conic shape is in the coordinate $X = 1/\Phi $ for the direction
transverse to the D3-brane. We can see the Dirac cone for a massless particle.

The agreement is not a coincidence. It is known that the free relativistic
electron with mass $m$ given by the Hamiltonian (\ref{Hamil2d}) is characterized
by a topological number, the TKNN number \cite{Thouless:1982zz}.
The TKNN number is defined as follows. First, we consider an eigenstate
of the Hamiltonian (\ref{Hamil2d}), 
\begin{eqnarray}
{\cal H}(p_1,p_2,m) \; v = \epsilon \; v \, 
\end{eqnarray}
which needs to be normalized, $v^\dagger v = 1$.
Using this eigenstate $v$, we define a Berry connection
\begin{eqnarray}
A_\mu^{\rm (B)} = v^\dagger i\frac{d}{dp_\mu} v \; (\mu=1,2)
\label{Berrymono}
\end{eqnarray}
Then this connection exhibits a topological property; the field strength
defined by the Berry phase shows half-integer quantization
\begin{eqnarray}
 \frac{1}{2\pi}\int\! dp_1 dp_2 \;  F_{12}^{\rm (B)}
  = \frac12 \text{sign}(m) \, .
  \label{signm}
\end{eqnarray}
In particular, taking a difference for positive and negative $m$, we obtain
the integral topological number 
\begin{eqnarray}
 \nu
  &\equiv& 
\frac{1}{2\pi}\int\! dp_1 dp_2 \;  
F_{12}^{\rm (B)}(m>0) - 
\frac{1}{2\pi}\int\! dp_1 dp_2 \;  F_{12}^{\rm (B)}(m<0) 
\nonumber \\
&=& 1 \, .
\label{TKNNnum}
\end{eqnarray}

At this stage we find a complete analogy with the Nahm construction of 
monopoles. The Berry connection (\ref{Berrymono}) is almost identical
to the formula for the gauge field in the Nahm construction of monopoles
(\ref{PhiAformula}). The difference is just the integral over $\xi$, which can be
shown to be irrelevant for the present case, due to the form of the eigenfunction
$\exp[r\xi]$.

Note that the space in which the monopole lives is $(p_1,p_2,m)$. 
Then the total monopole charge calculated by integrating the
magnetic flux surrounding the monopole located at the origin is
\begin{eqnarray}
1 = \frac{1}{2\pi}\int \! d{\bf S} \cdot {\bf B} = 
\frac{1}{2\pi}\int\! dp_1 dp_2 \;  
F_{12}\biggm|_{m>0} - 
\frac{1}{2\pi}\int\! dp_1 dp_2 \;  F_{12}\biggm|_{m<0} \, .
\end{eqnarray}
This is nothing but the calculation of the topological number (\ref{TKNNnum}).
So, we find that the monopole charge due to the Nahm construction
parallels the topological charge of the class A system.
Along the course, the band is identified with the shape of the D-brane.

\subsection{Generalization to topological number $\nu=2$}

\subsubsection{Copies of electrons for higher topological numbers}

The topological number calculated in the previous subsection is for
$\nu=1$. 
Generically, for the class A systems, the topological charge is a Chern
class labeled by integral topological number $\mathbb{Z}$, and it should
be possible to generalize it to the cases with more monopoles.%
\footnote{
Contrary to this, if the system is characterized by $\mathbb{Z}_2$
topological charge, it is not possible to make a situation with $\nu \ge
2$, which is just equivalent to $\nu=0$.
In this sense, the $\nu=2$ state plays an important role to distinguish
$\mathbb{Z}$ and $\mathbb{Z}_2$ systems.
}	
Here, we use the Nahm construction to obtain a fermion system which
has $\nu=2$.
We generalize the correspondence between the D-brane shape and the
electron dispersion to the case with $\nu=2$. 

As we have seen previously, the Nahm construction of monopoles coincides with
the Hamiltonian of a single free electron system.
To obtain $\nu=2$ 
we just follow the Nahm construction to find what kind of electron system is
relevant for the topological charge $\nu=2$. 

We have studied the $k=2$ monopole, and what we need to do is to look at
Nahm data and rephrase it to some electron Hamiltonians.
First, notice that Nahm's ``Dirac operator'' $\nabla^\dagger$ is a 
$2k\times 2k$ matrix for the monopole number $k$. The Nahm data $T_i$
are $k\times k$ matrices, which are tensored with the Pauli matrices $\sigma_i$.
So, to have a higher topological charge, we need to prepare $k$ copies of
electron Hamiltonians.

The Nahm data has to satisfy the Nahm equation (\ref{Nahm}). Since our 
extra dimension $\xi$ necessary for the Nahm construction corresponds to
an auxiliary parameter for the electron case, we look at $\xi$-independent
Nahm data. A generic solution to (\ref{Nahm}) is given by
\begin{eqnarray}
T_i = \Delta_i A
\label{TiA}
\end{eqnarray}
where $\Delta_i$ $(i=1,2,3)$ is an arbitrary 
real constant parameter, and $A$ is an arbitrary 2 by 2
Hermitian constant matrix. A special case was studied in (\ref{T1delta}). The Hermitian matrix can be decomposed to a unit matrix and Pauli matrices, so
\begin{eqnarray}
T_i = \Delta_i \left( {\bf 1}_2 + b_k \sigma_k\right)
\label{Tiallowed}
\end{eqnarray}
where $b_i$ $(i=1,2,3)$ are constant real parameters.

The corresponding electron system has a Hamiltonian
\begin{eqnarray}
{\cal H} = {\bf 1}_2\otimes(\sigma_1 p_1 + \sigma_2 p_2 + \sigma_3 m)  
- T_i \otimes \sigma_i 
\label{TKNN2}
\end{eqnarray}
The correspondence tells us that $T_i$ of the form (\ref{TiA})
exhibits the topological charge $\nu=2$. 
So we are led to a conclusion that the Hamiltonian (\ref{TKNN2})
is responsible for 2-dimensional topological insulator with topological charge 
$\nu=2$, 
once (\ref{TiA}) is satisfied. 

Note that the interacting Hamiltonian of the form (\ref{TKNN2})
with (\ref{TiA}) is a sufficient condition to get $\nu=2$. 
This is because we simply set $\xi$-independence for simplicity,
and generically various monopole solutions with $\xi$-dependent
Nahm data are possible.

\subsubsection{The shape of D-branes and electron dispersion for 
$\nu=2$} 

\begin{figure}[h]
\includegraphics[width=7cm]{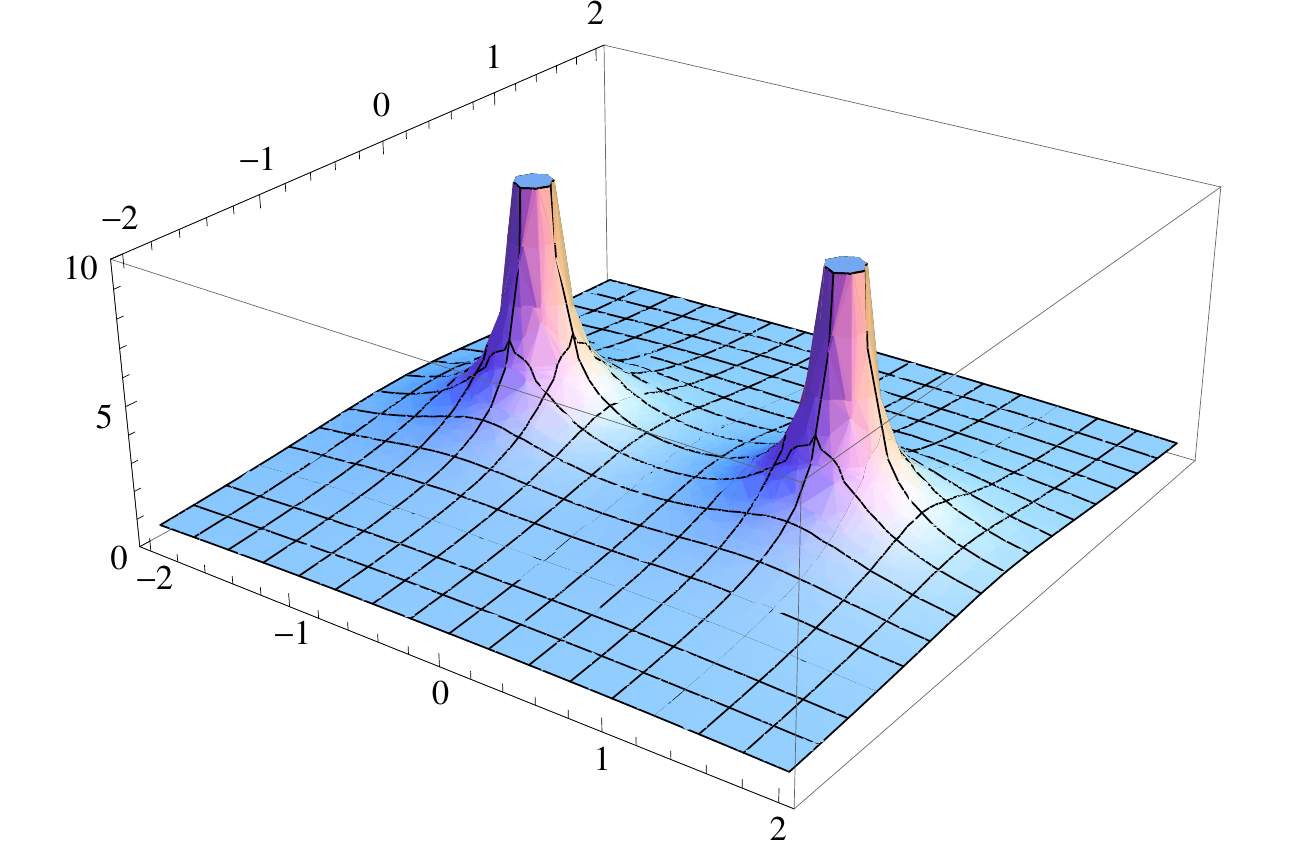}
\includegraphics[width=7cm]{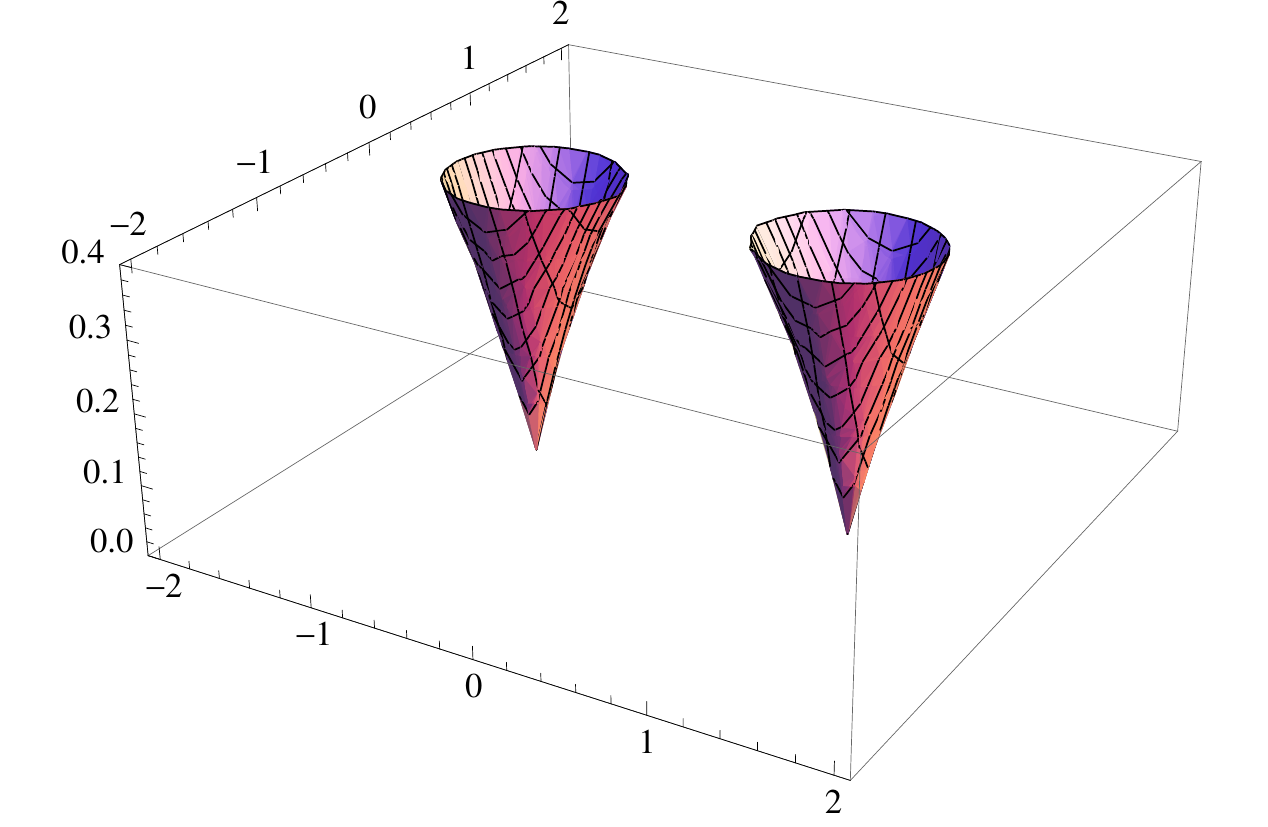}
\caption{(Left) The shape of the D3-brane for monopole number $k=2$.
The vertical axis is $-\Phi(x)$. (Right) A plot of $1/\Phi$. 
We find two Dirac cones.  }
\label{D3fig2}
\end{figure}

Here for simplicity we shall consider the simplest Nahm data (\ref{T1delta}) for the $k=2$ monopole in the Nahm construction. The resultant configuration of the D3-brane
consist of just two spikes whose centers are located at $x^1=\pm\Delta, x^2=x^3=0$. 
See the left figure in Fig.~\ref{D3fig2}. The two spikes correspond to two D1-branes
stuck to the D3-brane, and from the viewpoint of the worldvolume of the 
D3-brane, they are two monopoles.

Let us see a relation to the electron dispersion relation. The Hamiltonian
of the electron, corresponding to (\ref{T1delta}), is
\begin{eqnarray}
{\cal H}
=
\left(
\begin{array}{cccc}
m & p^1-ip^2 & & \Delta \\
p^1+ip^2 & -m^3 & \Delta & \\
& \Delta & m & p^1-ip^2 \\
\Delta & & p^1+ip^2 & -m^3
\end{array}
\right) .
\end{eqnarray}
We find two eigenvectors corresponding to $v_-$ and $v_+$ in (\ref{v-v+}).
The eigenvalues are
\begin{eqnarray}
\epsilon_\pm = \sqrt{(p_1\pm\Delta)^2 + (p_2)^2 + m^2}\, ,
\end{eqnarray}
therefore the zero of the momentum $p_1$ is shifted by $\Delta$. Using the correspondence (\ref{xp}), we find approximately
\begin{eqnarray}
2\epsilon = \frac{-1}{\Phi}
\end{eqnarray}
near the location of the monopoles. This expression is the same as that of 
the single monopole case, (\ref{energyscalar}).\footnote{Note that our scalar 
field $\Phi$ is single valued while there are two energy dispersions $\epsilon_{\pm}$. 
The reason is that in the Nahm construction we add two contributions from
$v_-$ and $v_+$.}
Plotting $-1/\Phi$, we find two Dirac cones (see the right figure of 
Fig.~\ref{D3fig2}).

\subsubsection{Bilayer graphene and Nahm equation}

For $m=0$, the Hamiltonian (\ref{TKNN2}) includes bilayer graphene. 
The monopole number $k=2$, which is the topological charge, needs
a twice large Hamiltonian matrix.
The enlarged Hamiltonian is naturally realized by a bilayer graphene.
The Nahm data term $\sigma_i \otimes T_i$ in (\ref{TKNN2}) naturally
encodes inter-layer interactions, for $b_1,b_2 \neq 0$.

Let us discuss briefly a correspondence to bilayer graphene
(see Ref.~\cite{CastroNeto:2009zz} for a  review of graphene). 
It is often convenient to describe the effective low energy Hamiltonian at
the continuum as
\begin{eqnarray}
{\cal H} =  
\left(
\begin{array}{cc}
H(p) & H_\perp \\
(H_\perp)^\dagger & H(p)
\end{array}
\right),\quad
H(p) \equiv
\left(
\begin{array}{cc}
0 & p^1-ip^2  \\
p^1+ip^2 & 0 
\end{array}
\right),
\end{eqnarray}
where $H_\perp$ is a $2\times 2$ generic matrix responsible for 
the inter-layer interactions. This $H_\perp$ can depend on the momentum $p$,
but here we consider only the case with $p$-independent constant matrix.

The first example is 
an AA-stacking bilayer graphene \cite{Liu:2009PRL} which 
has  the following inter-layer interaction
\begin{eqnarray}
H_\perp \equiv
\left(
\begin{array}{cc}
\Delta & 0  \\
0 & \Delta 
\end{array}
\right).
\end{eqnarray}
This corresponds to the following Nahm data
\begin{eqnarray}
T_1 = \Delta {\bf 1}_2, \quad T_2=T_3=0,
\end{eqnarray}
satisfying the Nahm equation (\ref{Nahm}). So it should naturally
exhibit $\nu=2$. 

On the other hand, an AB-stacking (Bernal stacking) bilayer graphene (see a review \cite{McCann:2013RPP}) may have
\begin{eqnarray}
 H_\perp
  = \left(
\begin{array}{cc}
  0&0  \\
\Delta   &0 
\end{array}
\right)
\end{eqnarray}
which is not consistent with the form (\ref{TiA}). Therefore our naive ansatz of having
a $\xi$-independent Nahm data does not work, so we cannot make sure that it 
has $\nu=2$. 

It would be an interesting question if the twisted graphene having typically the following effective Hamiltonian
\cite{de_Gail:2011PRB}
\begin{eqnarray}
 H_\perp
  =  \Delta \left(
\begin{array}{cc}
  1&1  \\
1   &1 
\end{array}
\right) + \cdots
\label{twisted}
\end{eqnarray}
can be in our category of Nahm data. The result should depend on whether the
first term  in (\ref{twisted}) dominates against others.

This argument is straightforwardly generalized to multilayer graphenes.
Applying the same procedure, we can similarly obtain higher topological
number situations, depending on how to stack the layers.

\subsection{All possible Hamiltonians for higher topological charges}

Since we have shown the equivalence between the dispersion relation of the 
2-dimensional topological insulators and the shape of D-branes, we would like 
to use the relation to discuss what are all possible Hamiltonians which
have generic topological number. 

The shape of D-branes has an intuitive understanding of what kind of configurations are possible. As we have seen in Fig.~\ref{D3fig2}, an example of the two-monopole
configuration suggests that possible configurations are only by positions of the monopoles. First, let us concentrate on the case of $k=1$ and $k=2$.

As we have learnt in this section, the eigenvalues of the Hamiltonian can either
be positive or negative, and they are paired. The eigenvalues $\pm\epsilon$
corresponds to $\pm r$ in the Nahm construction. So, choosing one sign 
for the energy means to choose $+r$ and discard $-r$.
In the Nahm construction, this in fact means that we consider only 
monopoles in $U(1)$ gauge theory. In order to construct non-Abelian monopoles,
one needs both $+r$ and $-r$ and restrict the allowed region of $\xi$ to be
a finite period. In our case, we need only the positive eigenvalues
which means that we treat $U(1)$ monopoles and an Abelian Berry connection,
as we assume that generic energy eigenstates are not degenerate.

For $k=1$, the generic solution of the Nahm equation (\ref{Nahm}) is
\begin{eqnarray}
T_i = c_i
\end{eqnarray}
where $c_i$ $(i=1,2,3)$ is a constant parameter. Looking at the zero-mode equation 
(\ref{DiracNahm}), we see that these $c_i$ specify the location of the monopole, 
$(x^1,x^2,x^3)=(c_1,c_2,c_3)$. Using our dictionary (\ref{xp}), this translates
to the zeros of the eigenvalues in the momentum space and the mass:
\begin{eqnarray}
(p_1,p_2,m) = (c_1,c_2,c_3) \, .
\end{eqnarray}
This means that for the topological charge $\nu=1$, 
once the BPS equation (\ref{BPSeq}) 
is assumed for the application of the Nahm construction,
all possible Hamiltonian is just given by a simple translation in
$(p_1,p_2,m)$-space.%
\footnote{%
Abelian monopoles are singular and normally the notion of moduli parameters
is not well-defined. 
However, one can find that physical observables of any Abelian BPS
monopole is parameterized only by its location.
}

Next, let us consider the case $k=2$. Once we assume that the Nahm data
is independent of $\xi$ (which may be a natural assumption since electron Hamiltonian does not depend on time $t$, since $\xi$ is interpreted as imaginary $t$), generic
solution of the Nahm equation is given by (\ref{Tiallowed}), other than the
total shift $(c_i)$.
Using the
redundant symmetry $T_i \rightarrow U T_i U^\dagger$ with any unitary transformation $U \in SU(k)$, we can further bring (\ref{Tiallowed}) to a diagonal form 
\begin{eqnarray}
T_i = c_i {\bf 1}_2 + d_i \sigma_3 = \left(
\begin{array}{cc}
c+d & 0 \\ 0 & c-d
\end{array}
\right).
\end{eqnarray}
So, basically, the Nahm data is the position of the two monopoles.
The D-brane configuration shown in Fig.~\ref{D3fig2} turns out to be generic.
Possible Hamiltonians with $\nu=2$ 
are parameterized only by the positions
of the monopoles, once the BPS equation for monopoles (\ref{BPSeq}) is assumed. 

We can generalize this argument to arbitrary monopole charge $k$. 
$\xi$-independent Nahm data $T_i$ means that the Nahm equation reduces to
\begin{eqnarray}
[T_i, T_j]=0\, .
\end{eqnarray}
The unique solution of this equation is made by diagonal matrices,
\begin{eqnarray}
T_i = U {\rm diag}(c_i^{(1)}, \ldots, c_i^{(k)}) U^\dagger
\end{eqnarray}
where $U$ is a generic $U(k)$ unitary matrix. So, generic monopole
configuration consists of just arbitrary distribution of center locations of
the $k$ monopoles. 

This confirms that the intuitive picture of D-branes exhaust all
possible Hamiltonians of 2-dimensional class A systems without any
additional symmetries.
Possible Hamiltonians having the nontrivial topological charge
$\nu=k$ 
is only dictated by the location of the $k$ monopoles in the $(p_1,p_2,m)$-space,
under the assumption that the monopoles
obey the BPS equations%
\footnote{The BPS equation (\ref{BPSeq}) for the $U(1)$ case
means that the gauge configuration satisfies Maxwell equation in vacuum, because
$\partial_i F_{ij} = \partial_i (\epsilon_{ijk} \partial_k \Phi) =
0$.
One may wonder why this needs to be satisfied for a generic Berry
connection.  The reason is simple: normally the Maxwell equation 
requires a current for a generic gauge connection ($\partial_i F_{ij} =
j_j$), but our Hamiltonian generically can have a rotation invariance in
$(p_1,p_2)$ space, so we can deduce $j_j=0$.
} and the electrons are free except for inter-layer momentum-independent interactions.

 \section{4D class A, dyonic instanton and D0-F1-D4 systems}
\label{sec3}

In this section, we consider a 4-dimensional class A systems
and study its relation to the shape of D-branes in string theory.
According to the periodic table presented in Table~\ref{PT}, this system
has integral topological charge $\bZ$, and hypothetical topological
insulators in 4 spatial dimensions are in fact classified by a second
Chern class, e.g. 4-dimensional quantum Hall
effect~\cite{Zhang:2001xs,Qi:2008ew}.
The Chern class counts Yang-Mills instanton number,
where generic solutions to self-dual equation of Yang-Mills are given 
by ADHM constructions \cite{Atiyah:1978ri,Corrigan:1983sv}. Furthermore, the instantons can be
regarded as a D-brane bound states: D0-brane sitting and dissolved 
inside the worldvolume of multiple D4-branes can be seen as a
Yang-Mills instanton configuration \cite{Witten:1995gx,Douglas:1995bn}. 
The shape of the
instanton can be detected again by introducing a scalar field, 
to form a dyonic instanton \cite{Lambert:1999ua}. This scalar field is nothing but the
shape of the D-brane corresponding to the dyonic instantons, which is
known \cite{Kim:2003gj} 
to be a D0-F1-D4 bound state and supertubes 
\cite{Mateos:2001qs,Emparan:2001ux,Mateos:2001pi} suspended
between parallel D4-branes.\footnote{In particular for the shape of the
dyonic instantons, see discussions in Ref.~\cite{Chen:2006rr}.}

In the following, first we briefly review the ADHM construction of instantons,
then study the shape of the corresponding D-branes to relate it to
electron dispersion in 4-dimensional topological insulators.


\subsection{A brief review of ADHM construction of instantons}

The ADHM construction of instantons 
\cite{Atiyah:1978ri,Corrigan:1983sv}\footnote{For the D-brane derivation
of the ADHM construction, see Ref.~\cite{Hashimoto:2005qh} and also 
Refs.~\cite{Akhmedov:2000zp,Akhmedov:2001jq}. }
allows calculating all solutions of
the self-dual equation of $SU(N)$ Yang-Mills theory,
\begin{eqnarray}
F_{\mu\nu} = * F_{\mu\nu} 
\end{eqnarray}
where $F_{\mu\nu} \equiv \partial_\mu A_\nu - \partial_\nu A_\mu -i [A_\mu,A_\nu]$
and the Hodge dual $*$ is given by $*F_{\mu\nu} = (1/2)\epsilon_{\mu\nu\rho\sigma} F_{\rho\sigma}$ ($\mu,\nu = 1,2,3,4$). 
The instantons are classified by the topological charge, namely the second Chern class,
as
\begin{eqnarray}
k = \frac{1}{16\pi^2}\int \! d^4x \;
{\rm tr} \left[
F_{\mu\nu} *F_{\mu\nu}\right] \, . 
\label{ADHMk}
\end{eqnarray}

The ADHM procedure to obtain the solutions with 
the instanton number $k$ is
as follows. 
First, we prepare ``Dirac operator''
\begin{eqnarray}
\nabla^\dagger \equiv
\left(
S^\dagger \;\;\; \bar{e}_\mu\otimes (x_\mu {\bf 1}_k-T_\mu)\right)
\end{eqnarray}
where $T_\mu$ ($\mu=1,2,3,4$) is a $k\times k$ Hermitian matrix,
and $S^\dagger$ is a $2k \times N$ complex matrix. 
The Pauli matrices are
$e_\mu=(i\sigma_1,i\sigma_2,i\sigma_3,{\bf 1}_2)$ and $\bar{e}_\mu$
is a complex conjugate of $e_\mu$.
The ADHM equation
which the ADHM data $S$ and $T_\mu$ need to satisfy is
\begin{eqnarray}
{\rm tr}\left[
\sigma_i \left(S^\dagger S + \bar{e}_\mu e_\nu T_\mu T_\nu\right)
\right]
=0 \quad (i=1,2,3)
\label{ADHMeq}
\end{eqnarray}
To construct instanton configurations, we solve the zero mode equation
\begin{eqnarray}
\nabla^\dagger v = 0
\end{eqnarray}
where $v$ is a vector with $N+2k$ components. Due to the size of $\nabla^\dagger$,
there exist $N$ independent vectors, so we label them as $v^{(a)}$ $(a=1,2,\ldots,N)$.
Then the gauge connection of the instanton, as a function of $x_\mu$, is given by
\begin{eqnarray}
A_\mu^{ab} \equiv 
i \left(v^{(a)}\right)^\dagger \frac{d}{dx_\mu} v^{(b)}.
\label{ADHMA}
\end{eqnarray}

Let us demonstrate how the simplest nontrivial case works, for our later purpose.
It is for $SU(2)$ Yang-Mills instanton with the instanton number $k=1$.
For $k=1$, $T_\mu$ in the ADHM data is just constant parameters, whose meaning is just a translation of $x_\mu$. So we can put $T_\mu=0$ without losing generality.
Then the ADHM equation (\ref{ADHMeq}) with generic $2\times 2$ complex matrix
$S = \alpha_\mu e_\mu$ provides
\begin{eqnarray}
{\rm tr}\left[
\sigma_i \bar{e}_\mu e_\nu 
\right]
\alpha^*_\mu \alpha_\nu
=0 \, . 
\end{eqnarray}
This condition amounts to 
$\alpha^*_k \alpha_l \epsilon_{ikl} + 
\alpha^*_4 \alpha_i  -\alpha^*_i \alpha_4  =0$. A generic solution to this equation 
is parameterized as $S = \rho U $ where $U$ is an $U(2)$ matrix and $\rho$ is
a complex constant parameter. Since this 
$U$ rotation does not change the final form of the gauge connection $A_\mu(x)$ of
the instantons, we can use $U$ transformation to simplify the ADHM data.  
Then we can take $S = \rho {\bf 1}_2$ with $\rho$ being a non-negative real parameter. The ``Dirac operator'' for the present case is
a $2\times 4$ matrix, 
\begin{eqnarray}
\nabla^\dagger = \left(
\rho {\bf 1}_2 \;\; \bar{e}_\mu x_\mu
\right) \, .
\label{Dirac4}
\end{eqnarray}

The normalized zero-mode $\nabla^\dagger v=0$ is solved as
\begin{eqnarray}
v = \frac{1}{\sqrt{r^2 + \rho^2}}
\left(
\begin{array}{c}
\bar{e}_\mu x_\mu\\
-\rho {\bf 1}_2
\end{array}
\right)
\label{ADHM1}
\end{eqnarray}
In this expression, two zero-modes are aligned to form a $4\times 2$ matrix $v$,
and we defined $r \equiv \sqrt{x_\mu x_\mu}$ as the distance from the center of the instanton in the 4-dimensional space. Using this zero-mode, following (\ref{ADHMA}),
we can calculate the
self-dual connection
\begin{eqnarray}
A_\mu=i v^\dagger \frac{d}{dx_\mu}v\, 
= \frac{i}{r^2+\rho^2} \left(e_\nu x_\nu \bar{e}_\mu - x_\mu {\bf 1}_2\right)
\label{Asingle}
\end{eqnarray}
With this, we can explicitly show
that the field strength is self-dual,
\begin{eqnarray}
F_{\mu\nu} = \frac12 \epsilon_{\mu\nu\rho\sigma} F_{\rho\sigma}
=\frac{i \rho^2}{(r^2+\rho^2)^2} \left(
e_\mu \bar{e}_\nu-e_\nu \bar{e}_\mu
\right),
\label{f=ftilde}
\end{eqnarray}
and 
the instanton number (\ref{ADHMk}) is $k=1$. 

The parameter $\rho$ is the size of the instanton. We will use the following property later,
\begin{eqnarray}
iv^\dagger \frac{d}{d\rho}v=0 
\label{rhovec}
\end{eqnarray}
for $v$ of this $SU(2)$ single instanton.

\subsection{4-dimensional class A system and ADHM construction}

Let us consider a free class A system in 4 spatial dimensions, whose
Hamiltonian is provided by
\begin{eqnarray}
{\cal H} = \gamma_\mu p_\mu + \gamma_5 m
\label{4dtopH}
\end{eqnarray}
where $\mu=1,2,3,4$ is for the four spatial directions, and $m$ is the mass. 
We have defined the gamma matrices
\begin{eqnarray}
\gamma_\mu \equiv \left(
\begin{array}{cc}
0 &\bar{e}_\mu \\ e_\mu & 0
\end{array}
\right), \quad
\gamma_5 \equiv -\gamma_1 \gamma_2\gamma_3\gamma_4 
=
\left(
\begin{array}{cc}
{\bf 1}_2 &0 \\ 0& -{\bf 1}_2
\end{array}
\right)
\end{eqnarray}
which satisfy the Clifford algebra $\{\gamma_M, \gamma_N\} =2 \delta_{MN} {\bf 1}_4$ ($M,N=1,2,3,4,5$).
The Hamiltonian eigenvectors satisfy
\begin{eqnarray}
{\cal H} v = \epsilon v
\end{eqnarray}
which is equivalent to
\begin{eqnarray}
\left(
\begin{array}{cc}
m-\epsilon &\bar{e}_\mu p_\mu \\ e_\mu p_\mu  & -m-\epsilon
\end{array}
\right) v = 0 \, .
\label{Ham4v}
\end{eqnarray}
Two normalized zero-modes are found as
\begin{eqnarray}
v = \frac{1}{\sqrt{(p_\mu)^2 + m^2}}
\left(
\begin{array}{c}
\bar{e}_\mu p_\mu \\
(\epsilon-m){\bf 1}_2
\end{array}
\right)
\label{Ham4vzero}
\end{eqnarray}
for which we need the dispersion relation
\begin{eqnarray}
\epsilon = \pm \sqrt{(p_\mu)^2 + m^2} \, .
\label{dis4}
\end{eqnarray}
The Berry connection is defined, using the zero-modes (\ref{Ham4vzero}), as
\begin{eqnarray}
A_\mu^{\rm (B)} = i v^\dagger \frac{d}{dp_\mu} v.
\label{Berry4}
\end{eqnarray}

At this stage, the analogy to the ADHM construction is obvious. 
The ``Dirac operator'' $\nabla^\dagger$ of the ADHM construction (\ref{Dirac4})
is nothing but the upper half part of
the matrix giving the zero-mode equation (\ref{Ham4v}), under the following identification:
\begin{eqnarray}
(x_\mu, \rho) \; \leftrightarrow \; (p_\mu, m-\epsilon)
\label{id4}
\end{eqnarray}

Then the ADHM connection (\ref{Asingle}) can be identified with the Berry connection
(\ref{Berry4}). Here the formulas look the same, but we need to be careful. In (\ref{Asingle}), the derivative $d/dx_\mu$ does not act on $\rho$, while in (\ref{Berry4})
the derivative $d/dp_\mu$ acts on $\epsilon(p)$ due to the dispersion relation
coming from the lower half of the Hamiltonian (which is absent in the ADHM construction). If we write the difference more explicitly, the Berry connection (\ref{Berry4}) is
\begin{eqnarray}
A_\mu^{\rm (B)} = i v^\dagger \frac{\partial}{\partial p_\mu} v
+  i \left(\frac{\partial\epsilon}{\partial p_\mu }\right)
v^\dagger \frac{\partial}{\partial \epsilon} v.
\end{eqnarray}
The second term is the difference from the ADHM construction.
However, interestingly, this difference vanishes for the single instanton case,
due to the special relation (\ref{rhovec}). So we conclude that Berry connection
(\ref{Berry4}) is identical to the ADHM connection (\ref{Asingle}) for the single instanton
in $SU(2)$.

Once the equivalence of the connection is given, one would think that 
the second Chern class should be equal to each other. Unfortunately, this
is not the case: the second Chern class differs from each other.
The reason is as follows. Since the instanton size $\rho$ corresponds to
$m-\epsilon(p)$, the size of the ``instanton'' of the 4-dimensional Hamiltonian
is not constant. It goes to zero at $p \rightarrow 0$, while it blows up at
$p \rightarrow \infty$. Therefore, although the second Chern class is
obviously topological, the value of the second Chern class for the 4-dimensional 
fermion system is not equal to
just $k=1$. 
This is also the case in 2 dimensions, as shown in
section~\ref{sec2}.
This half-integer quantization reflects the parity anomaly of the Dirac
fermion in odd dimensions, while its change must be integer.
See Ref.~\cite{Oshikawa:1994PRB} for the explanation in condensed-matter terminology.

This difference is again hidden in the derivative $\partial \epsilon/\partial p_\mu$.
Even though the BPST instanton connection (\ref{Asingle}) and the Berry connection
(\ref{Berry4}) are equal to each other, the field strengths are different, due to
$\partial \epsilon/\partial p_\mu$. In fact, for the Berry connection, we obtain
\begin{eqnarray}
F_{\mu\nu}^{\rm (B)} 
&=&\frac{i (m-\epsilon)^2}{(p^2+(m-\epsilon)^2)^2} \left(
e_\mu \bar{e}_\nu-e_\nu \bar{e}_\mu
\right)
\nonumber \\
& & +
\frac{2i}{(p^2 + (m-\epsilon)^2)^2}\frac{m-\epsilon}{p}\frac{\partial \epsilon}{\partial p}
e_\rho p_\rho 
\left(
p_\mu \bar{e}_\nu - p_\nu \bar{e}_\mu
\right),
\end{eqnarray}
which is {\it not self-dual}. The first line is equivalent to (\ref{f=ftilde}), 
while the second line
breaks the self-duality.
The second Chern class calculated with this field strength is provided as
\begin{eqnarray}
\frac{1}{16\pi^2} \int \! d^4p \, {\rm tr}
\left[
F_{\mu\nu}^{\rm (B)} * F_{\mu\nu}^{\rm (B)}
\right] = -\frac12 \mbox{sign}(m).
\end{eqnarray}
The second Chern class is a half integer, and 
we find an analogy to the 2-dimensional case, (\ref{signm}).
In particular, the second Chern class depend on the sign of the mass $m$,
and we find the difference for the change of the sign of the mass is
an integer,
\begin{eqnarray}
k\biggm|_{m>0} - k\biggm|_{m<0} = -1.
\end{eqnarray}

An interesting question is how we can obtain the Hamiltonian with 
the instanton number $k= 2$. The ADHM construction tells us
that the easiest way to get the multiple number of instantons is to have
a multiple system of ``sheets'' as in the case of graphene layers. If we tensor
the Hamiltonian (\ref{4dtopH}) such that the total Hamiltonian is
\begin{eqnarray}
{\cal H}^{\rm tot} = {\cal H} \otimes {\bf 1}_k
\end{eqnarray}
then this automatically has the instanton number $k$. The issue is what kind of
``inter-layer'' interaction does not spoil the topological number.
The total Hamiltonian can have off-diagonal interactions which are of the form
of the ADHM data $S$ and $T_\mu$. Once the ADHM data satisfy the ADHM equation
(\ref{ADHMeq}), the resultant connection satisfies the self-dual equation
and the instanton number remains $k$. 
However, our field strength of the Berry connection differs from that
of the instanton connection, as we have seen for the $k=1$ example.
So it is still an open question if these generic $k$ instanton ADHM data
corresponds to a larger topological number for the 4 dimensional class A
topological insulators.

\subsection{The shape of D-brane relates to band spectrum}

Our idea is to relate the shape of D-branes in the $x$ space 
to the 
band structure in the momentum $p$ space, via the identification
(\ref{id4}). The shape of D-brane is given by the transverse scalar field
living on the D-brane. 

We have seen above that the instanton charge dictates the 4-dimensional topological
insulators, and in string theory 
the instanton charge of $SU(2)$ Yang-Mills connection is
nothing but the D0-brane charge in 2 D4-branes \cite{Witten:1995gx,Douglas:1995bn}.
However, The D0-brane in the D4-branes does not modify the shape of 
the D4-brane itself, on the contrary to the case of the monopole where
the stuck D1-brane deforms the shape of the D3-brane so that we could identify
the shape of the deformed D3-brane given by the scalar field with the electron dispersion relation for the 2-dimensional systems.

There is a way to introduce a scalar field to the system of Yang-Mills 
theory; the dyonic instantons \cite{Lambert:1999ua}, namely, 
the instantons in the Coulomb phase. The scalar field appearing in the
dyonic instantons have a brane interpretation: it is indeed the shape of the
deformed D4-brane. The 2 D4-branes are separated parallelly from each other,
and the D0-brane needs to connect them with the help of fundamental strings (F1).
So, between the parallel D4-branes, there appears F1's with the D0-brane.
The configuration was first studied in Ref.~\cite{Kim:2003gj} and it was noticed
that the fundamental strings can blow up to form a supertube
\cite{Mateos:2001qs,Emparan:2001ux,Mateos:2001pi} suspended between
the parallel D4-branes. The supertube is a bound state of 
a cylindrical D2-brane with the fundamental strings and the D0-brane.

In the following, first we introduce the dyonic instantons and consider its
scalar field configuration, which is nothing but the shape of the D4-brane.
Then we will show that the shape can be identified with the electron dispersion
(\ref{dis4}) of the 4-dimensional topological insulator. The formula we will find
is similar to
the case of the 2-dimensional topological insulator, (\ref{energyscalar}).

The dyonic instanton is a solution to the following BPS equations 
in $(1+4)$-dimensional Yang-Mills-Scalar theory,
\begin{eqnarray}
F_{\mu\nu} = * F_{\mu\nu} , \quad
F_{\mu 0} = D_\mu \Phi  \, .
\end{eqnarray}
Here the subscript $0$ means the additional 
time direction, and $\Phi$ is the  scalar
field in the adjoint representation. The BPS equations solve the full equations 
of motion. In particular, the second equation can be solved by
simply setting $A_0 = \Phi$ with a Gauss law
\begin{eqnarray}
D_\mu D_\mu \Phi = 0.
\label{LaplacePhi}
\end{eqnarray}
It means that for $\Phi$ one needs to solve the Laplace equation
in the background of the Yang-Mills instantons.
Notice that the presence of $\Phi$ and the time component does not
modify the Yang-Mills instanton itself. The instanton configuration is
given first, then in that background one solves the Laplace equation (\ref{LaplacePhi}). 
So the information of the instanton parameters is not altered even when we 
upgrade the instanton to the dyonic instanton.

For our purpose we concentrate on $SU(2)$ Yang-Mills theory with a single instanton.
It is encouraging that there exists a formula to solve (\ref{LaplacePhi}) in the ADHM construction \cite{Dorey:2002ik},
\begin{eqnarray}
\Phi = \frac{C}{2} v^\dagger M v
\end{eqnarray}
where $M = {\rm diag} (1,-1,0,0)$. Here $C$ specifies the value of the scalar field
at the asymptotic infinity, 
\begin{eqnarray}
\Phi \to \frac{C}{2} \sigma_3 \quad (r \rightarrow \infty)
\label{asympPhi}
\end{eqnarray}
Using our zero-mode (\ref{ADHM1}) in the ADHM construction, 
after an appropriate unitary transformation, 
we obtain the scalar field of the dyonic instanton,
\begin{eqnarray}
\Phi = \frac{C}{2}\sigma_3 \frac{r^2}{r^2 + \rho^2} \, .
\label{D4shape}
\end{eqnarray}
This is nothing but the D4-brane shape deformed by the presence of the
D0-brane and the fundamental string.

\begin{figure}[t]
\includegraphics[width=7cm]{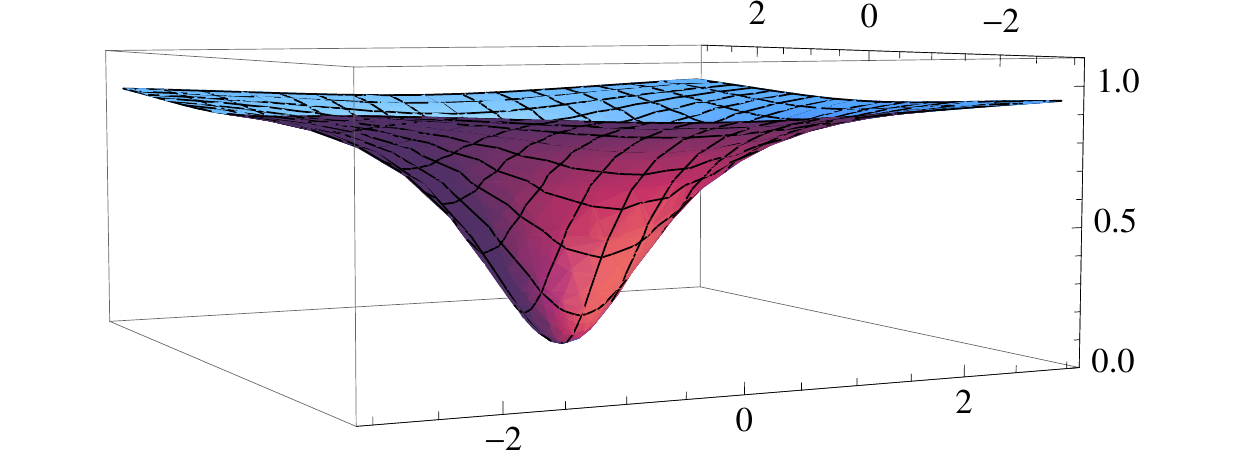}
\includegraphics[width=7cm]{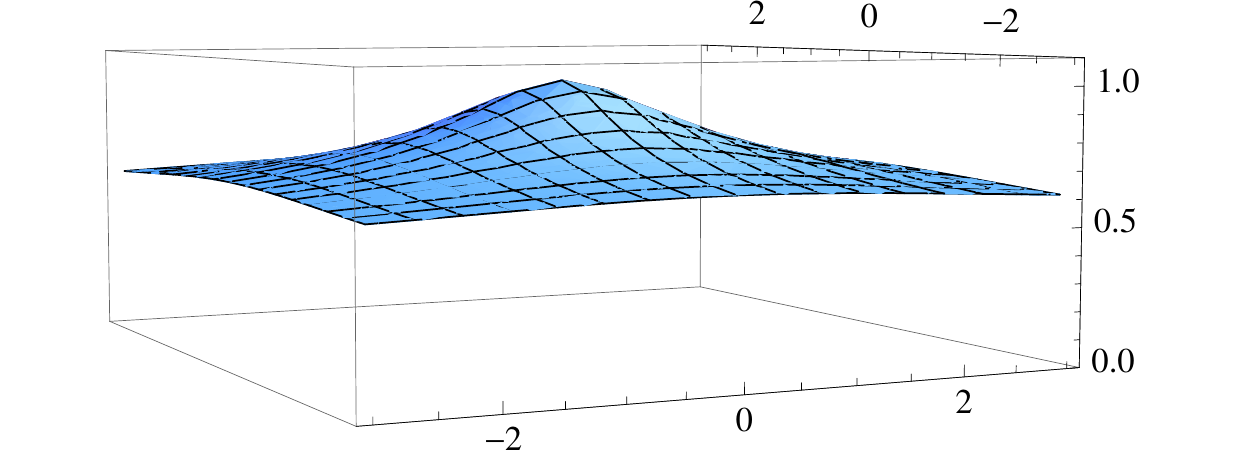}\\
\begin{center}
\includegraphics[width=7cm]{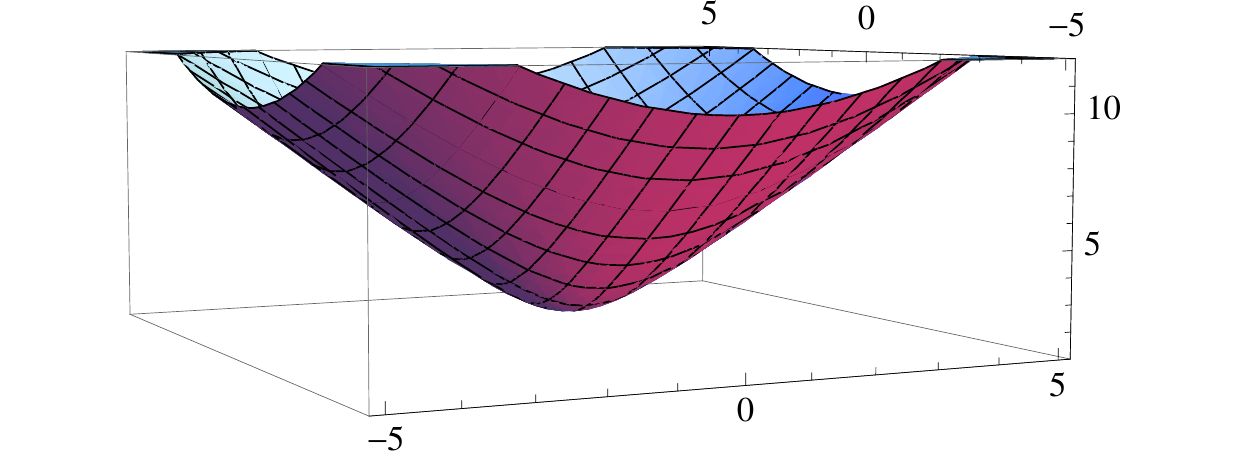}
\end{center}
\caption{Upper left: the shape $\Phi(x)$ of 
the D4-brane for a single dyonic instanton. The instanton size is
taken to be $\rho=1$, with $C=1$.  Upper right:
the shape $\Phi(p)$ for which the instanton size is
given through the dispersion relation $\rho = m-\epsilon(p) $.
Low: The electron dispersion given by 
$\epsilon = (m/2)/(\Phi(p)-1/2)$. }
\label{dyonicfig}
\end{figure}

Now, let us use our dictionary (\ref{id4}) to relate the D4-brane shape (\ref{D4shape})
to the electron dispersion (\ref{dis4}). Substituting the dictionary (\ref{id4}) to
the first component (which represents a D4-brane among the pair)  of 
(\ref{D4shape}), and choosing $C=2$ for simplicity (a scaling can recover the $C$ dependence anytime), we find
\begin{eqnarray}
\Phi = \frac{(p_\mu)^2}{(p_\mu)^2 + (m-\epsilon(p))^2} \, .
\end{eqnarray}
Using (\ref{dis4}), we can eliminate the explicit $p_\mu$ dependence in
this equation to finally obtain
\begin{eqnarray}
\epsilon = \frac{m/2}{\Phi -1/2} \, .
\label{energyscalar2}
\end{eqnarray}
This is the relation between the D4-brane shape $\Phi$ and the electron dispersion
$\epsilon(p)$. See Fig.~\ref{dyonicfig} for graphical images.

Remember that for the 2-dimensional system, the relation
was found in (\ref{energyscalar}) where the dispersion energy is
given by the inverse of the scalar field. Here, we find the same equation; 
The energy is given by the inverse of the scalar field, and the inverse is measured
from the asymptotic value of $\Phi$, 
\begin{eqnarray}
\Phi \rightarrow \frac{C}{4} =\frac12 \quad (p_\mu \rightarrow \infty)
\end{eqnarray}
Note that this is not inconsistent with the previous (\ref{asympPhi}). When $p$
goes to $\infty$, $\epsilon(p)$ also scales. This means that through the identification
(\ref{id4}), one needs to scale $\rho$ simultaneously. 

In summary, we have found that the band $\epsilon(p)$ for the 4 dimensional 
class A system can be identified as the shape of a D-brane, under the exchange 
$x_\mu \leftrightarrow p_\mu$. The precise relation is given by (\ref{energyscalar2}) through
the ADHM construction of dyonic instantons.


\section{Chiral edge mode, noncommutative monopoles and tilted D-brane}
\label{sec4}

So far, we have studied bulk properties and dispersions of electrons for
2-dimensional and 4-dimensional class A systems.
On the other hand, the essential feature of topological materials is to have a surface
massless state.
In this section, we shall demonstrate that the dispersion of the chiral
edge state for the 2-dimensional class A topological insulators can be
understood again as the shape of a D-brane.
The corresponding D-brane is a D-string which is tilted due to the
spatial noncommutativity on the worldvolume of a D3-brane.

First we will give a brief review of the chiral edge state, then we turn to 
a review of the
Nahm construction of a monopole in a noncommutative space.
We will see a correspondence between the dispersion of the chiral edge
state and the shape of the D-brane corresponding to the noncommutative
monopoles.

\subsection{Edge state and noncommutativity}

\begin{figure}[t]
\begin{center}
 \includegraphics[width=7cm]{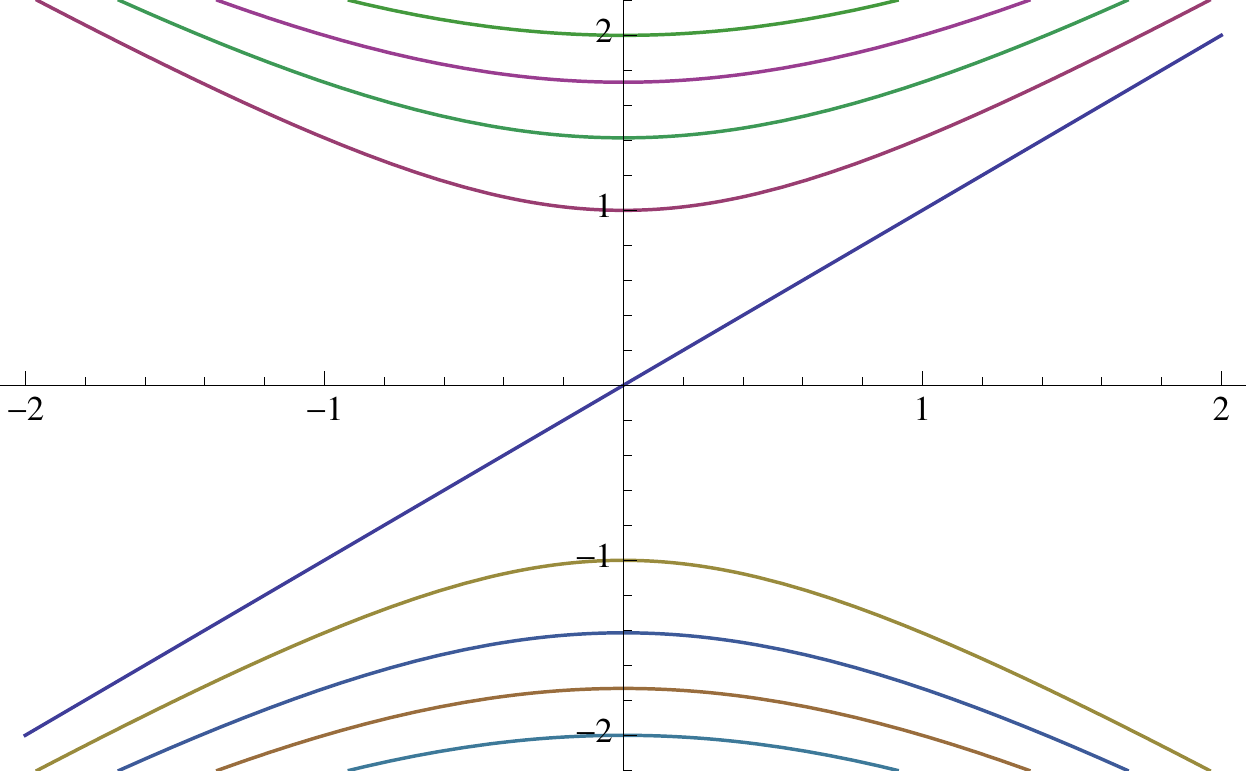}
\end{center}
\caption{The dispersion relation of the states at the edge with $\theta=1$.
The vertical axis is the energy $\epsilon$, and the horizontal axis is the
momentum $p_2$. There exists a straight line $\epsilon=p_2$ which
is the chiral edge state.}
\label{edgefig}
\end{figure}

First, we describe the chiral edge state appearing at a boundary of 
a 2-dimensional class A topological insulator, typically realized as the
quantum Hall effect (see Ref.~\cite{Volovik:2009univ} for a comprehensive review).
To introduce a boundary for the Hamiltonian (\ref{Hamil2d}) of the
2-dimensional system, let us consider an $x^1$-dependent mass term
$m(x^1)$, namely the domain-wall configuration.
$x^1=0$ is the boundary of the 2-dimensional material, where the gap closes.
To simplify the situation, we look at only the vicinity of the boundary,
and approximate the region by a linear profile of the mass, 
\begin{eqnarray}
m(x^1) = \theta x^1 \, .
\end{eqnarray}
At $x^1=0$, the mass changes its sign, which indicates the boundary
of the topological material such that the Chern number (\ref{TKNNnum})
is added once we cross the boundary line. The relevant Hamiltonian now reads\footnote{For our later purpose we exchanged the roles played by $\sigma_1$
and $\sigma_3$.} 
\begin{eqnarray}
{\cal H} = \sigma_1 \theta x^1 + \sigma_2 p_1 + \sigma_3 p_2. 
\end{eqnarray}
Due to the Heisenberg algebra $[\theta x^1, p_1] = i \theta$, we define
a creation/annihilation operator
\begin{eqnarray}
\hat{a} \equiv \frac{1}{\sqrt{2\theta}} \left(\theta x^1 + i p_1\right),
\quad
\hat{a}^\dagger  \equiv \frac{1}{\sqrt{2\theta}} \left(\theta x^1 - i p_1\right),
\end{eqnarray}
which satisfies $[\hat{a},\hat{a}^\dagger]=1$. 
(For simplicity, in this section we consider $\theta>0$.)
The Hamiltonian is conveniently written
as
\begin{eqnarray}
{\cal H} = \left(
\begin{array}{cc}
p_2 &\sqrt{2\theta} \; \hat{a}^\dagger \\
\sqrt{2\theta}\; \hat{a} & -p_2
\end{array}
\right).
\end{eqnarray}
Let us remark that this Hamiltonian is equivalent to the 2-dimensional
massive Dirac system in the presence of perpendicular magnetic field $B=\theta$, by
replacing the momentum $p_2$ with the mass.
In this sense, the dispersion relation shown in
Fig.~\ref{edgefig} is equivalent to the spectral flow with respect to the
mass parameter of the corresponding system.

The energy eigenstates with energy $\epsilon$ can be easily obtained as
\begin{eqnarray}
v_0 &=& \left(
\begin{array}{c}
|0\rangle \\ 0
\end{array}
\right), \quad \epsilon = p_2
\label{edge}
\\
v_n^{\pm} &=& 
{\cal N} \left(
\begin{array}{c}
\sqrt{2\theta n}|n\rangle \\ 
(\epsilon-p_2)|n-1\rangle
\end{array}
\right), \quad \epsilon = \pm \sqrt{2\theta n + (p_2)^2} \quad (n\geq 1)
\label{massiveA}
\end{eqnarray}
Note that the lowest mode is chiral, while higher modes are paired. The existence
of this chiral edge mode $\epsilon = p_2$ is related to the topological number 
$\nu=1$, 
which is called bulk-edge correspondence.
See, for example, a textbook on this topic~\cite{Wen:2004ym}.

The energy is parameterized by a continuous momentum $p_2$ and the
excited level $n$.
This is contrast to the bulk state of the 
2-dimensional system which is parameterized by the momenta 
$(p_1,p_2)$ and the mass $m$. Here, for the edge states, the noncommutativity
between $m$ and $p_1$ is important and discretize the bulk state as if
one has a magnetic field in the hypothetical $p_1$-$m$ plane.

If we look back our identification (\ref{xp}), we immediately notice
that the corresponding D-brane configuration should be through
a monopole in a non-commutative space, 
\begin{eqnarray}
[x^1,x^3]=-i\theta.
\end{eqnarray}
The monopole in a noncommutative space was first predicted by D-brane configurations in string theory in 
Ref.~\cite{Hashimoto:1999zw} and explicitly constructed in
Refs.~\cite{Gross:2000wc,Gross:2000ph,Gross:2000ss}.

\subsection{The shape of tilted D1-brane and chiral edge state}

The Nahm construction of monopoles in noncommutative space has been developed
in Refs.~\cite{Gross:2000wc,Gross:2000ph,Gross:2000ss} (see also Ref.~\cite{Hamanaka:2001dr}).
The construction is almost the same as that in a commutative space. There are two differences: first, $\hat{x}^1$ and $\hat{x}^3$ need to be treated as operators, obviously, and second, the Nahm equation (\ref{Nahm}) is modified to
\begin{eqnarray}
\frac{d}{d\xi}T_i = i\epsilon_{ijk}T_j T_k - \theta \delta_{i2}.
\end{eqnarray}
For a single $U(1)$ monopole, the simplest solution is
\begin{eqnarray}
T_2 = - \theta \xi, \quad T_1=T_3=0 \, .
\end{eqnarray}
Therefore the zero-mode equation of the ``Dirac operator'' is
\begin{eqnarray}
\left[
\frac{d}{d\xi} + \theta \xi \sigma_2 + \sigma_i x^i
\right] v=0 \, .
\end{eqnarray}
Note that here $\hat{x}^1$ and $\hat{x}^3$ are operators, and they do not commute.
The last term can be identified with the electron Hamiltonian at the edge.

A variation of the noncommutative monopole solutions include so-called ``fluxon''
solution \cite{Gross:2000ph,Polychronakos:2000zm}. It is a simple solution which is relevant to our study. The solution is given by the zero-mode
\begin{eqnarray}
v = \left(\frac{\pi}{\theta}\right)^{1/4}
\exp\left[-\frac{\theta}{2}\left(\xi + \frac{x_2}{\theta}\right)^2\right] \; 
\left(
\begin{array}{c}
|0\rangle\langle 0| \\ 0
\end{array}
\right)
\end{eqnarray}
One immediately notice a similarity to the chiral edge state (\ref{edge}).
As the region of $\xi$ for this zero-mode is given by $-\infty < \xi < \infty$,
we can evaluate the scalar field as
\begin{eqnarray}
\Phi = \int_{-\infty}^\infty d\xi \; v^\dagger \xi v
= -\frac{x_2}{\theta} |0\rangle\langle 0|
\label{phifluxon}
\end{eqnarray}
The scalar field configuration is linear in $x^2$. Indeed, the configuration
was interpreted in string theory as a D1-brane piercing a D3-brane 
at an angle given by the noncommutativity. When $\theta=0$, the D1-brane
becomes perpendicular to the D3-brane.

Since the Hamiltonian eigenvalues correspond to the operator $id/dt$
while the Nahm construction has $d/d\xi + \theta \xi$, from the formula
$\Phi = \int d\xi \; v^\dagger \xi v$ we expect $\theta \Phi \sim \epsilon$.
Indeed, 
comparing the D1-brane shape (\ref{phifluxon}) and the dispersion of
the chiral edge state (\ref{edge}), we find
\begin{eqnarray}
\epsilon = -\theta \Phi \biggm|_{|0\rangle\langle 0|}
\end{eqnarray}
under the identification (\ref{xp}). So, the shape of the piercing D1-brane
is the dispersion of the chiral edge state.


\section{2D class AII and D-branes with orientifold}
\label{sec5}

So far we studied class A systems which are fundamental examples
responsible for the quantum Hall effect.
Interesting topological insulators are offered with various other
classes as shown in Table \ref{PT}, and among them a popular topological
insulator is class AII exhibiting the time-reversal symmetry.
In this section we consider 2-dimensional class AII topological
insulators and provide a D-brane interpretation of the band spectrum.

\subsection{A brief review of helical edge state}

The class AII topological insulators are protected by a time-reversal symmetry.
A Hamiltonian of a free fermion which allows the time-reversal invariance can be
obtained by a combination of two class A Hamiltonians (\ref{Hamil2d}) 
which amounts to introducing the
spin degrees of freedom.
In fact, it is a Dirac Hamiltonian in 3 dimensions%
\footnote{
The model corresponds to the renowned Bernevig-Hughes-Zhang model \cite{Bernevig:2006Sci} for a topological
insulator. In our case 
the time-reversal invariant momentum is only at $p_1=p_2=0$ (since we work in
no lattice).
From the wave functions the topological number $\nu$ can be calculated as
$(-1)^\nu = - {\rm sign}(m)$ which means that the model has a nontrivial 
$\mathbb{Z}_2$ topological charge $\nu=1$ form $m>0$.
} 
with $p_3=0$, 
\begin{eqnarray}
{\cal H} = (\sigma_1 p_1 + \sigma_2 p_2) \otimes \sigma_2 + m \mathbf{1}_2 \otimes \sigma_3 \, .
\label{AIIH}
\end{eqnarray}
Since in this Dirac representation the spin operator is given by 
$(\hbar/2)\sigma_i \otimes \mathbf{1}_2$, 
the time-reversal symmetry transformation is provided as
\begin{eqnarray}
\Theta =  (-i\sigma_2) \otimes \mathbf{1}_2 K
\end{eqnarray}
where $K$ is the complex conjugation operator 
and the matrix $(-i\sigma_2)$ flips the sign of the spin operator. 
Under this operation, our Hamiltonian is transformed as follows,
\begin{eqnarray}
\Theta : \quad
{\cal H}(p) 
&\rightarrow &
\Theta^{-1} {\cal H}(p) \Theta
\nonumber \\
&&= 
\left[(-i\sigma_2) \otimes \mathbf{1}_2\right]^{-1} 
{\cal H}^*(p)
\left[(-i\sigma_2) \otimes \mathbf{1}_2\right] 
\nonumber 
\\
&&
= {\cal H}(-p) \, .
\label{TRI}
\end{eqnarray}
This shows the time-reversal symmetry for the Bloch Hamiltonian
for the present system, because 
$K$ changes the sign of the momenta $(p_1,p_2) \rightarrow (-p_1,-p_2)$
as is easily understood in the coordinate space representation.
Obviously from the definition we have $\Theta^2 = -1$,
which corresponds, in Table \ref{PT}, to the ``$-$'' sign in the class AII.

\begin{figure}[t]
\begin{center}
 \includegraphics[width=7cm]{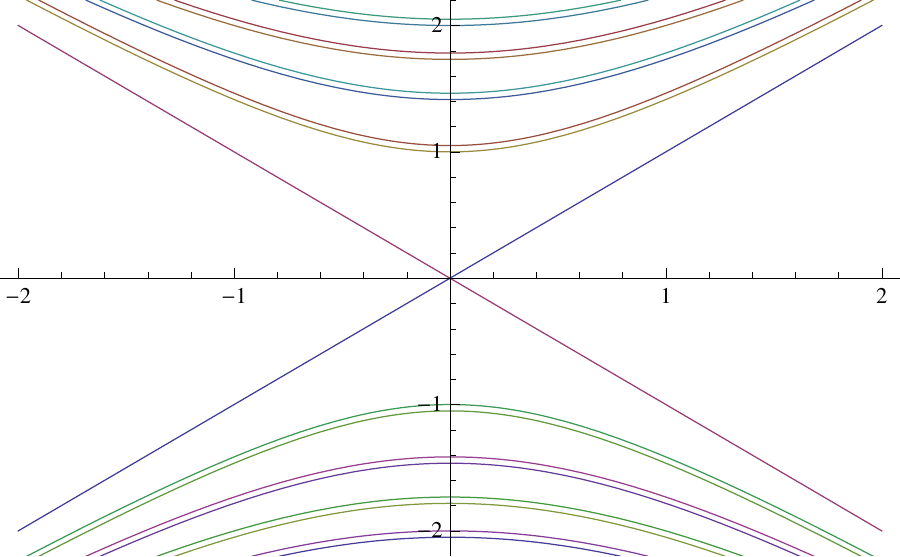}
\end{center}
\caption{The dispersion relation of the states at the edge.
The vertical axis is the energy $\epsilon$, and the horizontal axis is the
momentum $p_2$. There are two straight lines $\epsilon=\pm p_2$ which
are the helical edge states. Other gapped lines are doubly degenerate.}
\label{Helicalfig}
\end{figure}

To obtain the helical edge state, we repeat the procedures given in the last section.
First, consider the edge given by a mass profile $m(x^1) = \theta x^1$. 
Then, together with the momentum $p_1$, these form the creation and annihilation operators,
\begin{eqnarray}
m = \sqrt{\theta/2} (\hat{a} + \hat{a}^\dagger), 
\quad
p_1 = (-i)\sqrt{\theta/2} (\hat{a} - \hat{a}^\dagger). 
\end{eqnarray}
Hamiltonian eigenvalue problem can be easily worked out to have two
edge modes
\begin{eqnarray}
v_0^{(1)} = \left(
\begin{array}{c}
|0\rangle \\ 0 \\ 0 \\ i|0\rangle
\end{array}
\right), \quad \epsilon = p_2 \qquad
\mbox{and}
\qquad
v_0^{(2)} = \left(
\begin{array}{c}
0\\ |0\rangle \\ i|0\rangle \\ 0
\end{array}
\right), \quad \epsilon = -p_2 \qquad
\label{hedge}
\end{eqnarray}
These are helical edge modes, which are indicators of the class AII topological insulators.
The dispersion relations of the other modes (\ref{massiveA}) degenerate, see Fig.~\ref{Helicalfig}.

The class AII topological insulators have a topological number $\mathbb{Z}_2$. 
If one considers a deformation of the Hamiltonian (\ref{AIIH}) while 
keeping the time-reversal symmetry, the number of the pairs of the helical edge states
should be kept odd. That is, the intersection of the two straight lines in Fig.~\ref{Helicalfig}
cannot be reconnected to form a mass gap.


\subsection{Orientifold and helical edge state}

We would like to realize the spectra of the helical edge modes, $\epsilon = \pm p_2$,
in terms of the D-brane shape. As has been already constructed in the 
previous section, the dispersion of the chiral edge mode
corresponds to the shape of a slanted D1-brane. So, one would think that 
we just need to duplicate
the system such that we have two edge modes whose dispersion relations cross.
However, the story is not that simple. the most important property of the 
class AII topological insulators is the $\mathbb{Z}_2$ charge. As emphasized at
the end of the previous subsection, the helical edge modes should appear as
an odd number of pairs. How this property can be seen in the D-brane shape is
our goal of this section.

First, we need to re-interpret the Hamiltonian of the class AII, (\ref{AIIH}), in terms of
the Nahm construction. Renaming the variables $(p_1,p_2,m)$ to the coordinates 
$(x^1,x^2,x^3)$ as in (\ref{xp}), we can interpret the Hamiltonian as a Dirac operator
of the Nahm construction, by further adding $\partial/\partial \xi$.
At this stage, to make more use of the D-brane technique, we make use of the 
D-brane interpretation of the Dirac operator itself of the Nahm construction.
The interpretation was given in Ref.~\cite{Hashimoto:2005yy}: the essential interpretation
of the Dirac operator is a Hermitian tachyon field $T$ on a non-BPS D4-brane whose worldvolume
completely contains the D1 and the D3-branes,
\begin{eqnarray}
T= i\frac{\partial}{\partial \xi} + i u
\left[
(\sigma_1 x^1 + \sigma_2 x^2) \otimes (i\sigma_2) + \mathbf{1}_2 x^3  \otimes \sigma_3 
\right]
\end{eqnarray}
where $u$ is a real positive 
parameter which will be taken to infinity for the tachyon to be on-shell.
The four-dimensional space of the non-BPS D4-brane
worldvolume is spanned by the coordinates $(x^1,x^2,x^3,\xi)$.

\begin{figure}[t]
\begin{center}
\includegraphics[width=8cm]{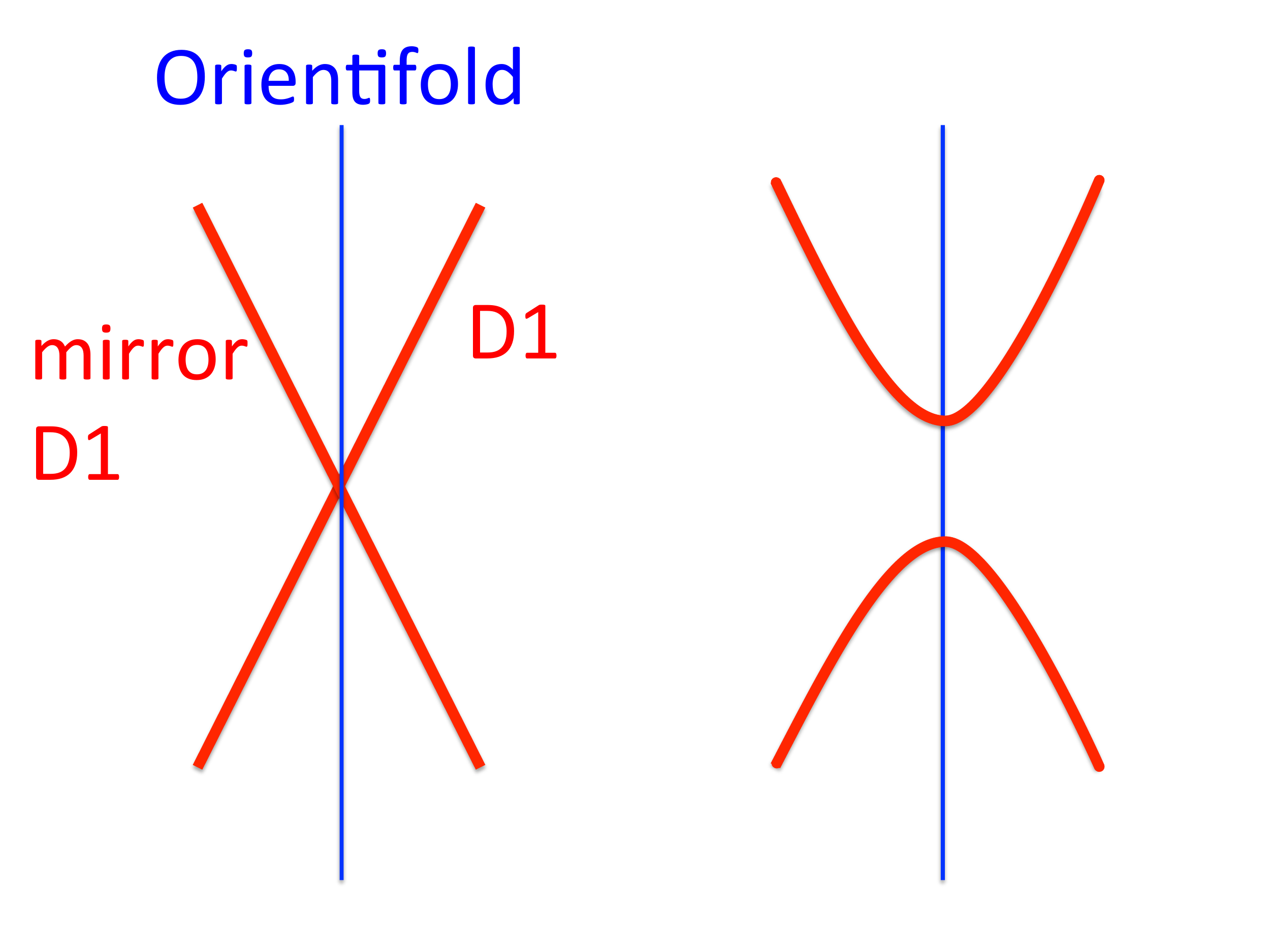}
\end{center}
\vspace{-10mm}
\caption{The effect of the orientifold in string theory. Left: a D1-brane intersects with the orientifold, yielding a mirror image of the D1-brane which appears to cross the original one at the orientifold.
Right: A prohibited D-brane configuration. The D-brane intersects with the orientifold cannot be a 
single D1-brane. Intersection on the orientifold should accompany another D1-brane.}
\label{orientifoldfig} 
\end{figure}

Now, we need to consider the time-reversal symmetry of the Hamiltonian. The D-brane 
interpretation of the time-reversal transformation should be
\begin{eqnarray}
\Theta : \quad
(x^1,x^2,x^3,\xi) \rightarrow (-x^1,-x^2,x^3,\xi), \quad 
T \rightarrow \left[(-i\sigma_2)\otimes \mathbf{1}_2\right]^{-1} 
T^*
 \left[(-i\sigma_2)\otimes \mathbf{1}_2\right]
\, .
\end{eqnarray}
The equivalence to the time-reversal transformation for the Hamiltonian (\ref{TRI}) is obvious.

Interestingly, this discrete transformation is equivalent to an orientifold transformation
acting on the D4-brane. The orientifolding in string theory is given by 
\begin{eqnarray}
T \rightarrow J^{-1} T^{\rm T} J
\end{eqnarray}
accompanied by a target space parity. Since the tachyon field (and the Dirac Hamiltonian) 
is Hermitian, we have $T^*= T^{\rm T}$. 
The orientifolding $J$ is for defining a symplectic structure acting on the
Chan-Paton factor of the non-BPS D4-branes, and our spin flip operation 
$(-i\sigma_2)\otimes \mathbf{1}_2$ for the fermion Hamiltonian is identified with $J$. 
Therefore, the time-reversal invariance means the existence of an orientifold localized at 
$x^1=x^2=0$ in string theory.

Now, the consequence of the existence of the orientifold is important. It directly shows how 
D-brane configurations are consistent with band spectra of topological insulators (see Fig.~\ref{orientifoldfig}).
\begin{itemize}
\item 
{\it Existence of a mirror D-brane.}

Suppose we have a slanted D1-brane as in the previous section, $\epsilon = p_2$. 
Then the orientifold shows the existence of another D1-brane which is a mirror partner 
of the original D1-brane, $\epsilon = -p_2$. So the D1-branes cross each other and the intersection 
is on the orientifold. 

This shape of the D-branes is nothing but the pair of the helical edge states.
\item
{\it D-brane intersect with the orientifold only as a pair.}

Upon the existence of the orientifold, it is known that D-branes crossing the orientifold
fixed plane need to move as a pair~\cite{Witten:1995gx,Gimon:1996rq}. 
In other words, a single D1-brane intersecting with the orientifold
is prohibited. Therefore the D1-brane intersecting with its mirror image cannot be reconnected.\footnote{The property comes from that of D5-branes in Type I superstring theory, where 
two coincident D5-branes have a $USp(2)$ gauge theory while they allow only a single scalar
field, showing that two D5-branes move together as a unit \cite{Gimon:1996rq}.}

This means that the helical edge state is stable against any deformation preserving the
time-reversal symmetry, and there is no way to produce a mass gap.
\end{itemize}
In summary, since the time-reversal invariance is identified with the existence of the orientifold
in string theory under the dictionary (\ref{xp}), the dispersion relation of the 
helical edge state is interpreted as a D1-brane intersecting with its mirror image on the orientifold.
No opening of a mass gap is consistent with the fact that D-branes needs to move as a pair  
on the orientifold.

\section{Summary and discussions}

In this paper, we showed that some dispersion relations and electron
band spectrum of class A topological insulators are the shape of
D-branes in string theory.
The former is in a momentum space while the latter is in a coordinate space.
The explicit dictionary between the momentum and the coordinate spaces
is given as (\ref{xp}) and (\ref{id4}), and the relations between the
electron bands and the shape of D-branes are given as (\ref{energyscalar}) and 
(\ref{energyscalar2}).
These examples are 2-dimensional and 4-dimensional class A topological
insulators, corresponding to quantum Hall effects.

The correspondence was found through an analogy of
the ADHM/Nahm construction of instantons and monopoles to the electron 
Hamiltonians of the topological insulators. The shape of D-branes is
captured by a scalar field on the D-brane, and the scalar field detects
the configuration of the topological solitons whose charge specify the 
topological properties of the electron system.

The correspondence between the electron band structure and the
shape of D-branes was further generalized to the chiral and helical edge
states.
The corresponding D-brane configuration represents a monopole
in non-commutative space which have been studied in string theory in details.
We found that fluxon solutions (\ref{phifluxon}) corresponds to the edge
state, and indeed the shape of the D1-brane piercing the D3-brane
corresponds to the dispersion of the chiral edge state.
This interpretation is also possible for the system with time-reversal
symmetry
giving a helical edge state. On the D-brane side the time-reversal
transformation is identified as an orientifolding, providing a mirror
image of the slanted D1-brane, resulting in a crossing of the D1-branes
at the orientifold.
Opening a mass-gap is not allowed because the crossed D1-branes cannot
be reconnected on the orientifold.

The intriguing part of the story is that D-brane picture is so intuitive
that it enables us to study generalization of the system.
As an example we studied the case with general integer value for
the topological number $\nu$. 
It turns out that under the assumption of the BPS
equation for the monopole and also under the assumption that the
inter-layer interaction does not depend on momentum, all possible
Hamiltonians having the general integer value $k$ of $\nu$ 
are characterized solely by the location of the $k$ monopoles in the
space spanned by $(p_1,p_2,m)$.

In this paper we considered topological insulators
only in 2 and 4 spatial dimensions.
Obviously it would be interesting to further consider the case with 3
dimensions.
The topological charge is ${\mathbb Z}_2$ for class AII system and the
dispersion relation of the helical edge states will be given by
some D-branes with an orientifold.
In addition, in this manner, Hamiltonian systems with dimensions higher
than 4 can be treated.
Therefore, it would be interesting to discuss how to realize 
all the topological superconductors in the classification table along
the direction studied in this paper.

D-branes have a lot of applications in string theory, and their shape can 
have various types. Spherical D-branes \cite{Myers:1999ps} and conic 
D-branes \cite{Hashimoto:2015wpa}
appear in various context in string theory. 
General electron band structure can be compared with the shape of 
generic D-brane configurations. 
Intersecting D-branes reconnect as described by the worldvolume gauge theories
\cite{Hashimoto:2003xz}, while electron bands reconnect generically in 
parameter space. More similarities and classifications due to our correspondence
would be important for possible topological materials and also for string theory dynamics.


\subsection*{Acknowledgments}
K.~H.~would like to thank M.~Sato for valuable discussions.
T.~K.~is grateful to Institut des Hautes \'Etudes Scientifiques for
hospitality where a part of this work has been done.
The work of K.~H.~was supported in part by JSPS KAKENHI Grant Numbers 15H03658, 15K13483.
The work of T.~K.~was supported in part by JSPS KAKENHI Grant Number 13J04302.

%

\bibliographystyle{ytphys}
\bibliography{band}

\end{document}